\documentclass[aps,prl,10pt,twocolumn,superscriptaddress,longbibliography]{revtex4}

\usepackage{graphicx}
\usepackage{amssymb, amsmath,amsfonts}
\usepackage{color}
\usepackage{ulem}
\usepackage{bbm}
\usepackage{graphicx}
\usepackage{subfigure}
\usepackage{dcolumn}
\usepackage{mathtools}
\usepackage{pifont}
\usepackage[utf8]{inputenc}
\usepackage[T1]{fontenc}
\usepackage{bbm}
\def\I{\uppercase\expandafter{\romannumeral 1}}
\def\II{\uppercase\expandafter{\romannumeral 2}}
\def\III{{\uppercase\expandafter{\romannumeral 3}}}
\def\IV{{\uppercase\expandafter{\romannumeral 4}}}
\def\V{{\uppercase\expandafter{\romannumeral 5}}}
\def\VI{{\uppercase\expandafter{\romannumeral 6}}}
\def\VII{{\uppercase\expandafter{\romannumeral 7}}}
\def\VIII{{\uppercase\expandafter{\romannumeral 8}}}
\def\i{\lowercase\expandafter{\romannumeral 1}}
\def\ii{\lowercase\expandafter{\romannumeral 2}}
\def\iii{{\lowercase\expandafter{\romannumeral 3}}}
\def\iv{{\lowercase\expandafter{\romannumeral 4}}}
\def\v{{\lowercase\expandafter{\romannumeral 5}}}
\def\vi{{\lowercase\expandafter{\romannumeral 6}}}
\def\vii{{\lowercase\expandafter{\romannumeral 7}}}
\def\nn{\nonumber\\}

\def\angstrom{\mbox{\normalfont\AA}}
\def\nn{\nonumber\\}

\def\k{\mathbf{k}}
\def\G{\mathbf{G}}
\def\Q{\mathbf{Q}}

\def\kt{\widetilde{\mathbf{k}}}

\def\qt{\widetilde{\mathbf{q}}}
\begin{document}

\title{Correlated insulators, density wave states, and their nonlinear optical response in  magic-angle twisted bilayer graphene}

\author{Shihao Zhang}
\affiliation{School of Physical Science and Technology, ShanghaiTech University, Shanghai 200031, China}

\author{Xin Lu}
\affiliation{School of Physical Science and Technology, ShanghaiTech University, Shanghai 200031, China}
\affiliation{Laboratoire de Physique des Solides, Univ. Paris-Sud, Universit\'e Paris Saclay, CNRS, UMR 8502, F-91405 Orsay Cedex, France}

\author{Jianpeng Liu}
\email[]{liujp@shanghaitech.edu.cn}
\affiliation{School of Physical Science and Technology, ShanghaiTech University, Shanghai 200031, China}
\affiliation{ShanghaiTech laboratory for topological physics, ShanghaiTech University, Shanghai 200031, China}

\begin{abstract}

The  correlated insulator (CI) states and the recently discovered density wave (DW) states in magic-angle twisted bilayer graphene (TBG) have stimulated intense research interest. However, up to date, the nature of these ``featureless" correlated states with zero Chern numbers are still elusive, and are lack of characteristic experimental signature.  Thus, an experimental probe to identify the  characters of these featureless CI and DW states are urgently needed. In this work, we theoretically study the correlated insulators and density-wave states at different integer and fractional fillings of the flat bands in magic-angle TBG based on extended unrestricted Hartree-Fock calculations including the Coulomb screening effects from the remote bands.  We further investigate the nonlinear optical response of the various correlated states, and find that the nonlinear optical conductivities can be used to identify the nature of these CI and DW states at most of the fillings. Therefore, we propose that nonlinear optical response can serve as a promising experimental probe to unveil the nature of the CI and DW states observed in magic-angle TBG.  

\end{abstract}


\maketitle

Twisted bilayer graphene (TBG) system around the magic angle provides a promising platform to achieve various intriguing quantum phases \cite{balents-review-tbg,andrei-review-tbg} such as the correlated insulators \cite{cao-nature18-mott,efetov-nature19,tbg-stm-pasupathy19,tbg-stm-andrei19,tbg-stm-yazdani19, tbg-stm-caltech19, young-tbg-science19,efetov-nature20,young-tbg-np20,li-tbg-science21}, orbital magnetic and Chern-insulator states \cite{young-tbg-science19, sharpe-science-19, efetov-arxiv20,yazdani-tbg-chern-arxiv20,andrei-tbg-chern-nm21,efetov-tbg-chern-arxiv20,pablo-tbg-chern-arxiv21,jpliu-nrp21}, as well as unconventional superconductivity \cite{cao-nature18-supercond,dean-tbg-science19,marc-tbg-19, efetov-nature19,efetov-nature20,young-tbg-np20,li-tbg-science21,cao-tbg-nematic-science21}. Near the magic angle 1.05$\,^{\circ}$   \cite{macdonald-pnas11}, there are two low-energy flat bands per spin per valley which are associated with nontrivial topological properties\cite{song-tbg-prl19, yang-tbg-prx19,po-tbg-prb19, origin-magic-angle-prl19, jpliu-prb19}. As a result, the electron-electron Coulomb interactions prevail  kinetic energy, and the interplay between the strong Coulomb correlations and the nontrivial band topology 
give rise to  diverse correlated and topological states in this system \cite{ kang-tbg-prl19,Uchoa-ferroMott-prl,xie-tbg-2018, zaletel-tbg-2019,zaletel-tbg-hf-prx20,jpliu-tbghf-prb21,zhang-tbghf-arxiv20,hejazi-tbg-hf,kang-tbg-dmrg-prb20,kang-tbg-topomott,yang-tbg-arxiv20,meng-tbg-arxiv20,Bernevig-tbg3-arxiv20,Lian-tbg4-arxiv20,regnault-tbg-ed,zaletel-dmrg-prb20,macdonald-tbg-ed-arxiv21,meng-tbg-qmc-cpl21,lee-tbg-qmc-arxiv21,bultinck-tbg-strain-prl21}. 

Most of the previous works focus on the integer fillings, and only a few pioneering works have paid attention to the fractional fillings of the flat bands in TBG \cite{pablo-tbg-chern-arxiv21,young-tbmg-cdw-arxiv21,xie-tbgfci-nature21,philips-nanoletter-18,zaletel-tbg-kekule-PRX21,kim-tbg-arxiv21}, which may realize unconventional density-wave (DW) states \cite{pablo-tbg-chern-arxiv21,young-tbmg-cdw-arxiv21,xie-tbgfci-nature21} and even fractional Chern-insulator states \cite{xie-tbgfci-nature21}. However,  up to date the nature of most of the ``featureless" correlated insulator (CI) states observed at  integer fillings such as the zero-Chern-number CIs  at $\nu=3, 0, \pm2$ \cite{cao-nature18-mott,efetov-nature19}, and the recently observed DW states at  the fractional fillings $\nu\!=\!7/2$, and 11/3 \cite{xie-tbgfci-nature21}),  are still elusive. An experimental probe to distinguish the  characters of these  states is  needed.

In this work, we use an extended unrestricted Hartree-Fock (HF) method within the subspace of the flat bands 
to study CIs and DW states at  all integer fillings $-3\leq\nu\leq3$, and a few fractional fillings $\nu=8/3$, $7/2$, and 11/3, at which  CI and DW states are observed. The Coulomb potentials acted on the flat-band subspace from the occupied remote bands are taken into account \cite{Bernevig-tbg3-arxiv20}, and the screening of Coulomb interactions in the flat-band subspace are treated by constrained random phase approximation (cRPA). We further study different components of the nonlinear optical conductivities of all the symmetry-breaking states in the valley-sublattice space,   and propose that the various competing correlated states in TBG can be  identified through the nonlinear optical response.

\begin{figure}[!htbp]
\includegraphics[width=0.5\textwidth]{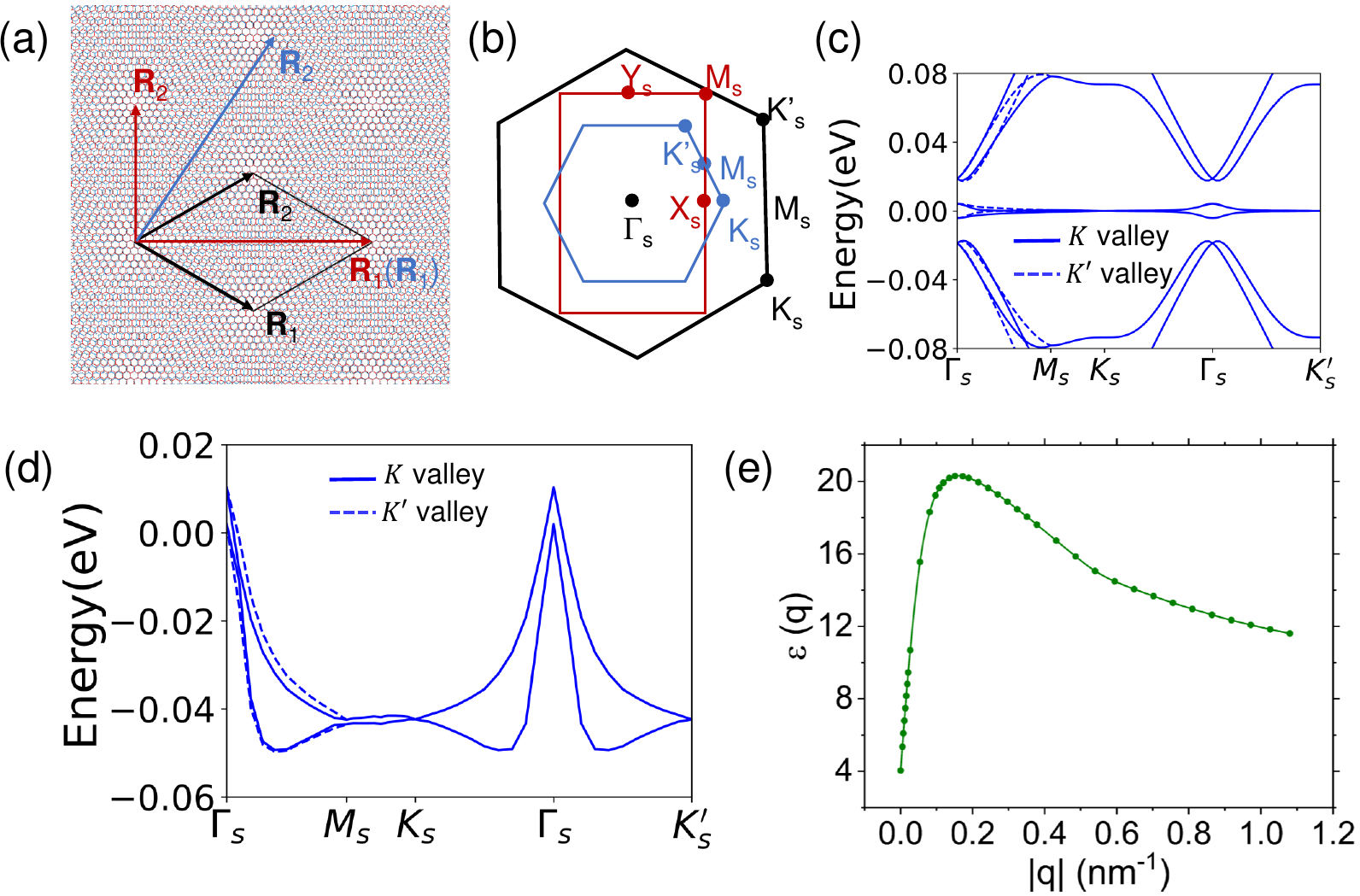}
\caption{~\label{fig1} (a) Illustration of the moir\'e superlattice of twisted bilayer graphene.  The lattice vectors of the primitive moir\'e cell, the doubled moir\'e supercell, and $\sqrt{3}\times\sqrt{3}$ moir\'e supercell are marked by black, red, and blue arrows, respectively. (b) Moir\'e Brillouin zones of the primitive cell, doubled supercell, and $\sqrt{3}\times\sqrt{3}$ tripled supercell. (c) The non-interacting energy bands of magic-angle TBG, and (d) the flat-band dispersions including remote-band Hartree-Fock potentials.
(e) The wavevector dependence of the effective dielectric constant calculated by cRPA method.}
\end{figure}

We first introduce the non-interacting Hamiltonian and the moir\'e superlattice geometries used in this work. In Fig.~\ref{fig1}(a), the primitive  moir\'e cell is marked by  black rhombus, and the real-space moir\'e lattice vectors of the primitive cell, the doubled supercell, and the $\sqrt{3}\times\sqrt{3}$ tripled supercell are marked by black, red, and blue vectors, respectively; the corresponding reciprocal lattice vectors  and moir\'e Brillouin zones of the three types of moir\'e supercells are shown in Fig.~\ref{fig1}(b). 
The low-energy effective Hamiltonian for TBG of valley $\mu$ ($\mu=\mp$ for $K/K'$ valley) is described by Bistritzer-MacDonald continuum model\cite{macdonald-pnas11}: 
\begin{equation}
H^{0}_{\mu}=\begin{pmatrix}
-\hbar v_{F}(\hat{\mathbf{k}}-\mathbf{K}^{\mu}_1)\cdot\mathbf{\sigma}_{\mu} & U_{\mu}(\mathbf{r}) \\
U^{\dagger}_{\mu}(\mathbf{r}) & -\hbar v_{F}(\mathbf{k}-\mathbf{K}^{\mu}_2)\cdot\mathbf{\sigma}_{\mu}
\end{pmatrix}\;,
\label{eq:Ham}
\end{equation}
where $v_F$ denotes the Fermi velocity, and the Hamiltonian is expanded near the Dirac points of the two layers $\mathbf{K}^{\mu}_1$ or $\mathbf{K}^{\mu}_2$,  with  $\mathbf{\sigma}_{\mu}=(\mu\sigma_x,\sigma_y,\sigma_z)$ ($\mu=\pm$) denoting Pauli matrices in the sublattice space. $U_{\mu}(\mathbf{r})$  refers to the interlayer coupling matrix \cite{supp_info}. In Fig.~\ref{fig1}(c), we show the energy bands of the continuum model of the $K$ valley at the magic angle $\theta=1.05^{\circ}$ in the moir\'e Brillouin zone (mBZ) of the primitive cell, which can be classified into two flat bands near the charge neutrality point (CNP) and remote bands above and below them.

Now we consider the dominant intra-valley component of the long-range Coulomb interactions in the TBG system \cite{supp_info}, 
\begin{equation}
H_C\!=\!\frac{1}{2N_s}\sum _{\lambda \lambda^{\prime}}\sum _{\mathbf{k} \mathbf{k^{\prime}}\mathbf{q}}\,V(\mathbf{q})\,\hat{c}^{\dagger}_{\mathbf{k+q},\lambda}\,\hat{c}^{\dagger}_{\mathbf{k^{\prime}-q}, \lambda^{\prime}}
\,\hat{c}_{\mathbf{k^{\prime}}, \lambda^{\prime}}\,\hat{c}_{\mathbf{k},\lambda}
\label{eq:coulomb}
\end{equation}
where $N_s$ denotes the total number of moir\'e cells in the system, $\mathbf{k}$ and $\mathbf{q}$ represent  wavevectors relative to the Dirac points, $\lambda\equiv(\mu,\alpha,\sigma)$ is a composite index, with $\mu$, $\alpha$, $\sigma$ referring to the valley, layer/sublattice, and spin indices, respectively. A double-gate screened Coulomb interaction, $V(\mathbf{q})\!=\!e^2\tanh(|\mathbf{q}|d_s)/(\,2\Omega _M \epsilon_{\textrm{BN}}\varepsilon_0 |\mathbf{q}|\,)$ is adopted, where $\Omega _M$ is the area of moir\'e supercell, $d_s=40\,$nm, 
$\epsilon_{\textrm{BN}}\approx 4$,  
and $\varepsilon_0$ is the vacuum permittivity. 
Then we self-consistently  solve the interacting Hamiltonian $H_0+H_C$ with unrestricted Hartree-Fock approximation, which will be extended to be adapted for the doubled and tripled moir\'e supercells. 

In Ref.~\onlinecite{Bernevig-tbg3-arxiv20} and Ref.~\onlinecite{kang-cascade-tbg-arxiv21}, the authors propose to ``regularize" the Coulomb interaction by subtracting a constant density $(1/2) \delta_{\mathbf{q},\mathbf{0}}$ from the density operator $\hat{\rho}(\mathbf{q})$, which reads 
\begin{equation}
H_{C}^{\prime}=\frac{N_s}{2}\sum_{\mathbf{q}}\,V(\mathbf{q})\,\delta \hat{\rho}(\mathbf{q})\,\delta \hat{\rho}(-\mathbf{q}),
\label{eq:HC}
\end{equation}
in which $\delta \hat{\rho}(\mathbf{q})$ is defined as the density matrix at wavevector $\mathbf{q}$ subtracted by a constant $(1/2)\,\delta_{\mathbf{q},\mathbf{0}}$\cite{supp_info}. 
Note that Eq.~(\ref{eq:HC}) is not normal ordered. After being projected onto a subset of low-energy bands, e.g., the flat bands,  Eq.~(\ref{eq:HC}) would differ from its normal ordered form $H_C$, and the difference   $\Delta H_C=H_{C}^{\prime}-H_{C}$ can be re-expressed as the HF potentials exerted by the remote bands on the flat bands by virtue of the particle-hole  and $C_{2z}\mathcal{T}$ symmetry of the projected interaction Hamiltonian \cite{Bernevig-tbg3-arxiv20}. In  Fig.~\ref{fig1}(d), we present the flat band structures  including remote-band HF potentials. Clearly, $\Delta H_{C}$ significantly enhances the overall bandwidth of flat bands and induces the particle-hole asymmetry. In addition to the screening effects from the metallic gates, the Coulomb interactions between electrons in the flat bands of TBG can be further screened by virtual particle-hole excitations from the remote bands, 
which are treated by cRPA \cite{pollet-tbg-crpa-prb20}, and the details are presented in Supplementary Information \cite{supp_info}.
The calculated effective dielectric constant as a function of wave vector $\mathbf{q}=\widetilde{\mathbf{q}}+\mathbf{Q}$ 
is shown in Fig.~\ref{fig1}(e), which is consistent with the previous report \cite{pollet-tbg-crpa-prb20}.

We  perform unrestricted HF calculations within the low-energy subspace of the flat bands including remote-band HF potentials, with the cRPA screened Coulomb interactions. We first study  the ground states at  integer fillings $-3\leq\nu\leq3$ without breaking moir\'e translational symmetry. 
With a realistic parameter  choice of the continuum model \cite{supp_info,koshino-prx18},
 our calculations reveal that the ground state at CNP is a Kramers intervalley coherent (K-IVC) state characterized by order parameters $(\tau_x,\tau_y)\sigma_y$ \cite{zaletel-tbg-hf-prx20}, with mixture of small valley polarization component $\vert\langle\tau_z\rangle\vert\!\approx\!0.1$ , where $\mathbf{\tau}$ and $\mathbf{\sigma}$ denote Pauli matrices defined in the valley and sublattice space, respectively.
The slight mixture of valley polarization into the KIC state at CNP results from the remote-band screening effects \cite{supp_info}.
At $\nu=\pm 1$, 
the HF ground state is also a mixed state with both K-IVC and spin-valley polarized (SVP) orders with Chern number $\pm 1$, consistent with previous theoretical and experimental reports \cite{Lian-tbg4-arxiv20,efetov-arxiv20,jpliu-tbghf-prb21}. 
At $\nu=\pm2$, with realistic parameter choice,  we find that the ground state is a fully spin and valley polarized (SVP) state based on Hartree-Fock+cRPA calculations.
 However, if we take a fixed dielectric constant $\epsilon=10$, then the ground state at $\nu=2$ involves the nearly degenerate IVC and SVP states \cite{supp_info}. This indicates that the actual ground state at $\nu=2$ is subtle, and can be sensitive to details of the system.  
At $\nu=\pm 3$, the calculated HF ground state is always a SVP state with Chern number $\pm1$.  In Table~\ref{table:integer} we present the dominant order parameters, symmetries,  valley polarizations, Chern numbers, and gaps of the HF ground states preserving moir\'e translational symmetry at all integer fillings. We find that most of these CI states exhibit significant valley polarizations, which will contribute to nonlinear optical responses as will be discussed below.  The band structures from the HF+cRPA calculations are given in Supplementary Information \cite{supp_info}.

\begin{table}[!t]
 \caption{Symmetries, order parameters, valley polarizations, Chern numbers $C$, and gaps for the ground states at integer fillings.}
 \label{unitcell}
 \centering
\begin{tabular}{cccccc} 
 \hline
 $\nu$ &  order parameters & symmetry & $| \langle \tau_z \rangle |$ & $|C|$ & gap (meV)\\ 
 \hline
 $3$ & $\tau_z,s_z,\tau_zs_z$ & $C_{3z},C_{2x}$ & 1 & 1   & 7.1 \\
 $2$ & $\tau_z,s_z,\tau_zs_z$ & $C_{3z},C_{2z}\mathcal{T},C_{2x}$ & 2 & 0 & 15.3\\
 $1$ & $(\tau_x,\tau_y)\sigma_y,\tau_z,s_z,\tau_zs_z$ & $C_{3z}$ & 0.85 & 1 & 7.5\\
 $0$ & $(\tau_x,\tau_y)\sigma_y,\tau_z$ & $C_{3z}$ & 0.10 & 0 & 25.0\\
 $-1$ & $(\tau_x,\tau_y)\sigma_y,\tau_z,s_z,\tau_zs_z$ & $C_{3z}$ & 1.14 & 1 & 6.1\\
 $-2$ & $\tau_z,s_z,\tau_zs_z$ & $C_{3z},C_{2z}\mathcal{T},C_{2x}$ & 2 & 0 & 14.1\\
 $-3$ & $\tau_z,s_z,\tau_zs_z$ & $C_{3z},C_{2x}$ & 1 & 1 & 4.4\\
 \hline
 \label{table:integer}
\end{tabular}
\end{table}

\begin{figure}[!htbp]
\includegraphics[width=0.5\textwidth]{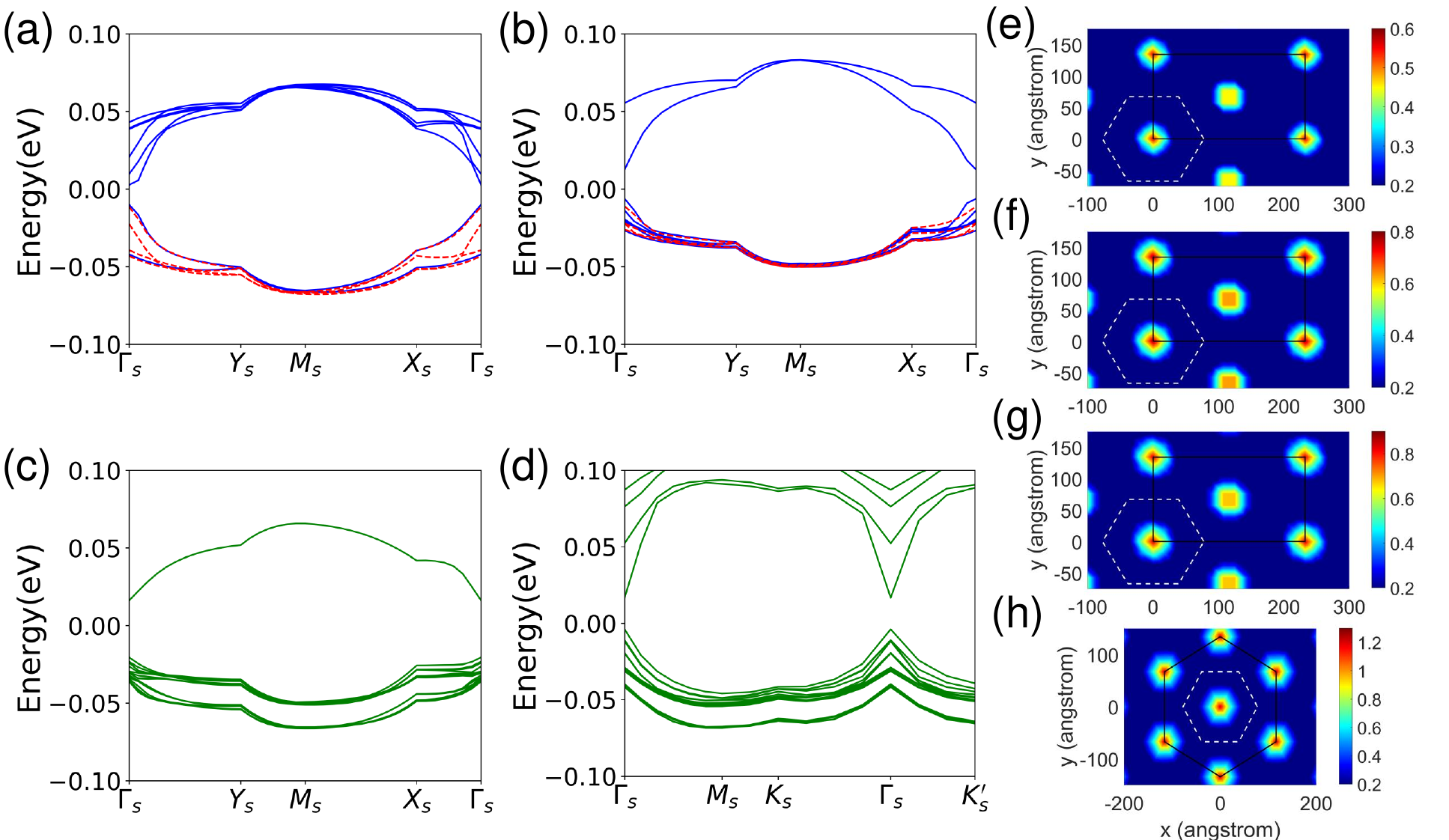}
\caption{~\label{fig3} The Hartree-Fock energy bands of density-wave states for (a) $\nu$\,=\,1 (b) $\nu$\,=\,3, (c) $\nu$\,=\,7/2, and (d) $\nu$\,=\,8/3. The real-space distributions of charge density for (e) $\nu$\,=\,1,  (f) $\nu$\,=\,3, (g) $\nu$\,=\,7/2, and (h) $\nu$\,=\,8/3. The  moir\'e primitive cell is marked with white dash lines.}
\end{figure}


CI state with zero Chern number has been observed at $\nu=3$ in TBG \cite{efetov-nature19}, which is inconsistent with previous theoretical results assuming preserved moir\'e translational symmetry \cite{zhang-tbghf-arxiv20,Lian-tbg4-arxiv20,hejazi-tbg-hf}. Thus one naturally expects that the zero Chern number state at $\nu=3$ may involve a spontaneous moir\'e translational symmetry breaking. Therefore, in this work we also perform HF calculations at $\nu=3$ based on a doubled moir\'e supercell (see Fig.~\ref{fig1}(a)-(b)).  Moreover, motivated  by the recent discoveries of correlated states at fractional fillings of the flat bands \cite{xie-tbgfci-nature21}, we also study the possible DW states at $\nu=1, 7/2,  8/3$ and 11/3. Let us first consider the  DW state at  $\nu\!=\!3$  in TBG with doubled moir\'e supercell. The system still stays in the SVP phase at $\nu\!=\!3$ , but there is one pair of unoccupied energy bands from the same valley-spin flavor due to the cell doubling, and both of them have zero Chern numbers.  
The real-space charge density distribution of this zero-Chern-number DW state is inhomogeneous around the neighbouring $AA$ sites, clearly breaks the primitive moir\'e translational symmetry as shown in Fig.~\ref{fig3}(f).  We also study the possible DW state with doubled moir\'e supercell at $\nu=1$,  and the ground state is  a SVP state with zero Chern number, with $C_{3}$-broken charge distributions as shown in Fig.~\ref{fig3}(e).  
The HF band structures of these DW states with doubled primitive cell are presented in Fig.~\ref{fig3}(a) ($\nu\!=\!1$) and (b) ($\nu\!=\!3$), where the solid blue  and red dashed lines denote bands from the $K$ and $K'$ valleys respectively. 

We continue to study the DW states at fractional fillings 7/2, 8/3, and 11/3, which are calculated based on doubled ($\nu=7/2$) and  $\sqrt{3}\times\sqrt{3}$ tripled ($\nu=8/3, 11/3$) moir\'e supercells, respectively. The choice of such supercells can be justified by generalized susceptibility calculations \cite{supp_info}. The calculated ground states at 7/2, 8/3, and 11/3 fillings are gapped, which may explain the experimentally observed $C\!=\!0$ correlated insulators at 7/2 and 11/3 fillings, and the unusual $C\!=\!1$ Chern insulator  at $\nu=8/3$, as reported in Ref.~\onlinecite{xie-tbgfci-nature21}. In particular, at 7/2 filling, we find two nearly degenerate ground states with the energy difference $\sim 10\,\mu$eV: one is a zero-Chern-number spin-polarized K-IVC state with slight valley polarization, and the other is a SVP state having Chern number 1. We propose that the experimentally observed $C\!=\!0$ insulator state at 7/2 is the spin polarized K-IVC state. 
With doubled primitive moir\'e cells, there are 8 flat bands from each valley, and the K-IVC order would mix the two valleys, opening a gap between the 8 valence flat bands and the 8 conduction flat bands, all with zero Chern numbers.  At 7/2 filling, 7 out of the 8 conduction flat bands are filled, yielding a zero-Chern-number spin polarized K-IVC  state.  The single-particle spectrum  around 7/2 filling is presented in Fig.~\ref{fig3}(c). 

At 8/3 filling,  based on $\sqrt{3}\times\sqrt{3}$ tripled supercell HF calculations, we find the ground state is  a SVP state with slight mixture of K-IVC order, and the calculated Chern number of this state is $1$ \cite{supp_info}, consistent with experiments \cite{xie-tbgfci-nature21}. Such a state is adiabatically connected to a pure SVP state with Chern number 1 \cite{supp_info}. At 11/3 filling, the ground state turns out to be a pure SVP state with zero Chern number. 
It should be noted that the bandwidth of the conduction flat band can be significantly reduced in the presence of vertical magnetic fields due to the orbital magnetic effects, concomitant with a more uniform distribution of Berry curvatures \cite{xie-tbgfci-nature21,parker-fqah-arxiv21}. As a result, around filling 11/3 the system would undergo a transition from the DW state to the fractional Chern insulator state with increased magnetic field \cite{xie-tbgfci-nature21,parker-fqah-arxiv21}. 

The Hartree-Fock band structures at 7/2 and 8/3 fillings are presented in Fig.~\ref{fig3}(c) and (d), respectively. The corresponding ground-state real-space charge density distributions are presented in Fig.~\ref{fig3}(g) ($\nu=7/2$) and (h) ($\nu=8/3$), respectively. 
The leading order parameters, symmetries, the calculated valley polarizations $\langle\tau_z\rangle$, and the calculated IVC order amplitudes  in the different DW states at $\nu=1, 3, 7/2, 8/3$ and 11/3 are presented in Table.~\ref{table:dw}.

\begin{table}[!t]
 \caption{Symmetries, order parameters,  valley polarizations $\langle\tau_z\rangle$, K-IVC order amplitudes and gaps for  the different DW states}
 \label{table:dw}
 \centering
\begin{tabular}{cccccc} 
 \hline
 $\nu$ & main order parameters & symmetry & $\vert\langle \tau_z \rangle\vert$ & IVC & gap (meV)\\ 
 \hline
 $1$ & $\tau_z,s_z,\tau_zs_z$ & $C_{2z}\mathcal{T}$ & 6 & 0 & 12.7\\
 $3$ & $\tau_z,s_z,\tau_zs_z$ & $C_{2z}\mathcal{T}$ & 2 & 0 & 17.2\\
 $7/2$ & $(\tau_x,\tau_y)\sigma_ys_{0,z},s_z$ & $\mathcal{T}^{\prime}$ & 0.07 & 0.45 & 35.7\\
 $8/3$ & $\tau_z,s_z,\tau_zs_z, (\tau_x,\tau_y)\sigma_y$ & $C_{3z}$ & 3.65 & 0.54 & 20.6\\
 $11/3$ & $\tau_z,s_z,\tau_zs_z$ & $C_{2z}\mathcal{T}, C_{3z}$ & 1 & 0 & 52.4\\
 \hline
 \label{table:dw}
\end{tabular}
\end{table}


Quite a few of the experimentally observed CIs and DWs  in magic-angle TBG are featureless insulators, such as the CIs observed at CNP, $\nu=\pm2$, the CI at $\nu=3$, and the DW states at $7/2$, etc.  All of  these states are just insulating with zero Chern number, and do not exhibit any particular signature in conventional transport and optical approaches. Therefore, it is difficult to experimentally identify the nature of these correlated states. Here we propose that these featureless correlated states may exhibit distinct nonlinear optical responses, which is described by the generation of an alternating current density $j^{c}(\omega_1+\omega_2)$ due to the second-order response to the electric fields:
\begin{equation}
j^{c}(\omega_1+\omega_2)=\sum_{a,b=x,y}\sigma^{c}_{ab}(\omega_1+\omega_2)E_a(\omega_1) E_b(\omega_2)\;.
\end{equation}
 The nonlinear optical conductivity tensor $\sigma^{c}_{ab}$ may serve as a promising probe to unveil the nature of the various correlated states observed in magic-angle TBG. In this work, we study two kinds of nonlinear optical processes, the shift-current response with $\omega_1=-\omega_2=\omega$, and the second hardmoic generation (SHG) with $\omega_1=\omega_2=\omega$.  For SHG response, the nonlinear susceptibility  $\chi^{c}_{ab}(2\omega)=i\sigma^{c}_{ab}(2\omega)/(\varepsilon_0 2\omega)$. 

We illustrate our idea by comparing the nonlinear optical responses of two prototypical CIs both with zero Chern numbers: a K-IVC state characterized by order parameter $(\tau_x\sigma_y, \tau_y\sigma_y)$, and a VP state characterized by order parameter $\tau_z$.  A VP order actually spontaneously breaks both $C_{2z}$, time-reversal ($\mathcal{T}$), and $C_{2y}$ symmetries, and preserves $C_{2z}\mathcal{T}$, $C_{3z}$, and $C_{2x}$ symmetries.  Such a state exhibits counter-propagating current loops in real space, which contribute to staggered orbital magnetic fluxes \cite{jpliu-tbghf-prb21}.  Since both  $C_{2z}$ and $\mathcal{T}$ symmetries are broken due to such real-space current pattern (although $C_{2z}\mathcal{T}$ is preserved), a VP state can have nonlinear optical response with symmetry-allowed nonlinear optical conductivity components: $\sigma_{xx}^{x}(\omega)=-\sigma_{xy}^{y}(\omega)=-\sigma_{yx}^{y}(\omega)=-\sigma^{x}_{yy}(\omega)$ \cite{supp_info}. In particular, a non-vanishing $\sigma_{xx}^{x}(\omega)$ component is a smoking gun for the valley polarization, i.e., $\sigma_{xx}^{x}(\omega)=\sigma_{xx,z}^{x}(\omega)\langle\tau_z\rangle$ \cite{supp_info}.  On the other hand, although a K-IVC order $(\tau_x\sigma_y, \tau_y\sigma_y)$ generally breaks $C_{2z}$ symmetry,  the combination of $C_{2z}$ symmetry and a $C_{2z}'=\tau_z C_{2z}$ symmetry, would enforce the vanishing nonlinear optical response of a K-IVC state \cite{supp_info}. Thus nonlinear optics can be considered as a reliable approach to distinguish the VP and K-IVC states. Further  analysis reveal that all the IVC states have vanishing nonlinear optical response, and there are only  three types of order parameters, i.e., the VP order $\tau_z$, the ``nematic order" $(\tau_z\sigma_x,\sigma_y)$, and the sublattice order $\sigma_z$, that are allowed to have nonzero nonlinear optical responses in TBG \cite{supp_info}. We note that the order parameters $(\tau_z\sigma_x,\sigma_y)$ are also involved in the ``incommensurate Kekul\'e state" which are proposed as the ground states at non-zero fillings of magic-angle TBG under finite strain \cite{zaletel-tbg-kekule-PRX21}.

\begin{figure}[!htbp]
\includegraphics[width=0.5\textwidth]{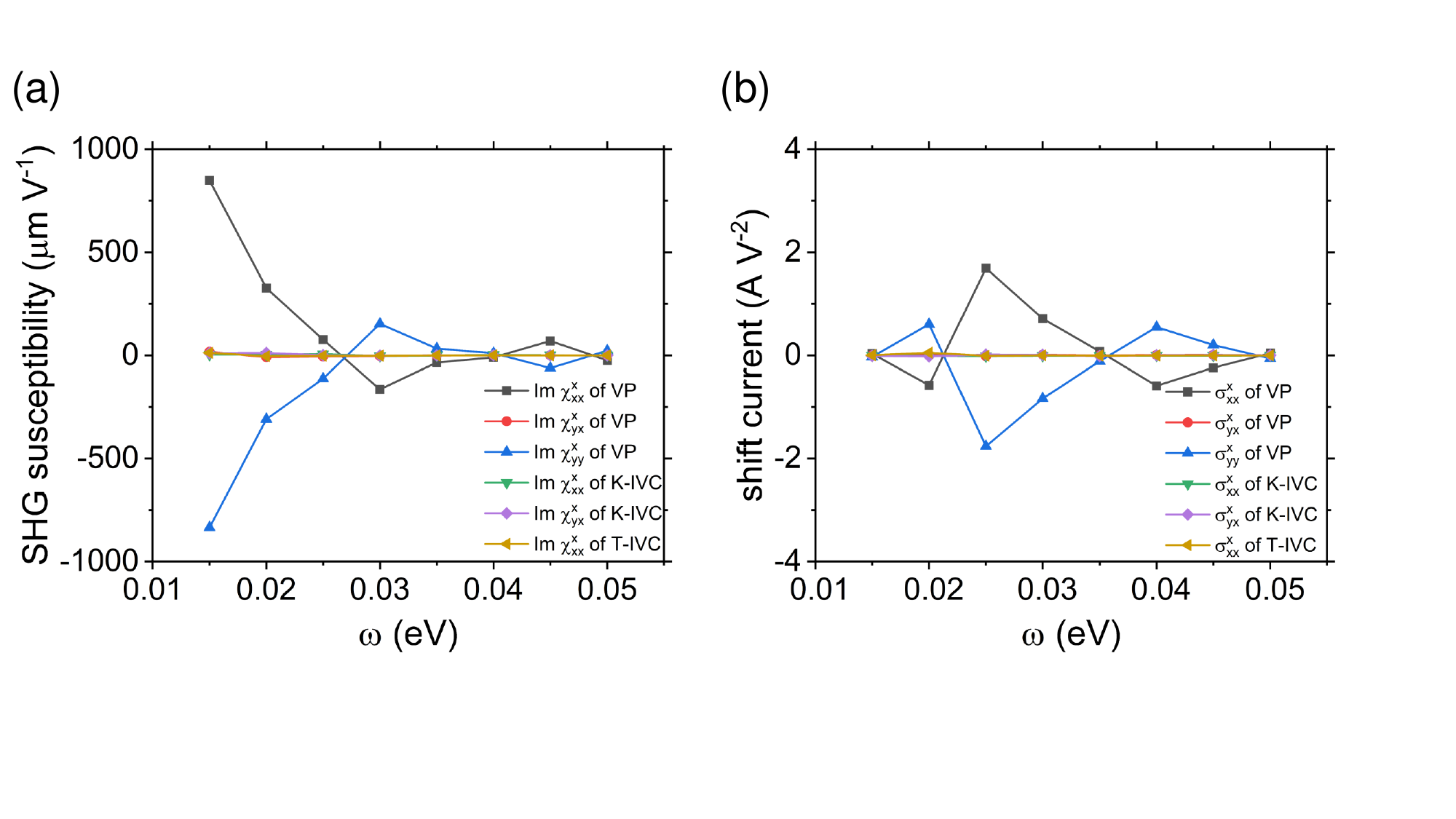}
\caption{~\label{fig4} The nonlinear optical response of TBG at the CNP assuming  constant order parameters with amplitudes (1\,meV) applied to the flat bands.}
\end{figure}

In order to verify the above argument, we have numerically calculated the SHG susceptibility and nonlinear photo conductivity for the shift current for various different ordered states in TBG, assuming a constant, $\mathbf{k}$ independent order parameter with amplitude of $1\,$meV in each state, with the filling fixed at CNP.  The results are presented in Fig.~\ref{fig4}(a) (for SHG) and (b) (for shift current). We see that the nonlinear susceptibilities vanish for both K-IVC state, and the time-reversal invariant intervalley coherent (T-IVC) state characterized by order parameter $(\tau_x\sigma_{x},\tau_y\sigma_{x})$. In contrast, the SHG susceptibility and shift-current conductivity of a VP state are giant in the infra-red frequency regime $\hbar\omega\sim 15$-25\,meV.

Now we discuss the feasibility of performing nonlinear optical measurements for a  magic-angle TBG device. Nowadays the size of a good TBG sample can be made more than 10$\,\mu$m in each lateral direction, whereas the spot size of a laser beam in the infrared frequency regime (say, with the wavelength of 2500\,nm) can be adjusted to  $\sim\!5\,\mu$m \cite{john-discussion,yao-shg-matter20},  smaller than the size of the TBG sample. Thus one can safely  rule out any undesirable response from the sample boundaries, and directly perform SHG rotation anisotropy measurements for a bottom-gated TBG device with vertical incident light. Actually SHG measurements have already been performed for non-magic-angle TBG device \cite{yao-shg-matter20}. By virtue of the chiral structure of TBG, remarkable second-harmonic signal contributed by the $\chi_{xyz}$ component of the SHG susceptibility tensor has been observed \cite{yao-shg-matter20}. This experiment indicates that performing SHG measurements on magic-angle TBG device is promising and feasible. Moreover, given  that a couple of ``hidden" correlated phases have already been unveiled by SHG measurements in some strongly correlated systems such as cuprates and iridates \cite{Torre-CuO-SHG-natphys2021,Zhao-iridate-SHG-natphys2016,harter-science17}, we would  expect the successful application of this technique to the magic-angle TBG system as well.

To summarize, in this work we have theoretically studied the correlated insulators and density wave states at various integer and fractional fillings of the flat bands based on extended unrestricted HF calculations with screening effects treated by cRPA method. We have explained  the nature of the recently observed density wave states and symmetry-breaking Chern insulator states at $\nu=1, 3$, $7/2$, and $8/3$ fillings.  We also find that most of the CI and DW states exhibit substantial valley polarizations, which would contribute to giant nonlinear optical response. We have identified the symmetry-allowed nonlinear optical conductivity components in different types of ordered states.
 Our results indicate that nonlinear optical response may serve as a promising probe to distinguish different types of correlated states observed in magic-angle TBG, which would stimulate further experimental and theoretical studies on the nonlinear optical properties of moir\'e 2D systems.  

\acknowledgements
\paragraph{Acknowledgements. \textemdash} This work is supported by the National Key R \& D program of China (grant no. 2020YFA0309601), the National Science Foundation of China (grant no. 12174257), and the start-up grant of ShanghaiTech University. We would like to thank P Xi Dai,  Jian Kang, John A. McGuire, and Chengjiang Du for valuable discussions.   We thank the HPC platform of ShanghaiTech University for providing the computational resource.
\textit{Note added}: We are aware of a few related works  posted on arXiv during the submission of our work: in Ref.~\onlinecite{wagner-tbg-iks}, the authors have extensively studied the incommensurate Kekul\'e state at non-integer fillings of TBG; in Refs. \onlinecite{Bernevig-TBG-STM-arxiv21,Zaletel-TBG-STM-arxiv21}, two theory groups independently proposed to distinguish the different correlated states in magic-angle TBG using scanning tunnelling microscopy. %

\bibliography{tmg}

\widetext
\clearpage

\begin{center}
\textbf{\large Supplementary Information for “Correlated insulators, density wave states, and their nonlinear optical response in  magic-angle twisted bilayer graphene"}
\end{center}

\vspace{12pt}
\begin{center}
\textbf{\large \I\ The continuum model for twisted bilayer graphene}
\end{center}
The Bistritzer-Macdonald continuum model is adopted to describe the low-energy physics of twisted bilayer graphene around the magic angle: 
\begin{equation}
H^{0}_{\mu}(\mathbf{r})=\begin{pmatrix}
-\hbar v_{F}(\hat{\mathbf{k}}-\mathbf{K}^{\mu}_{1})\cdot\mathbf{\sigma}_{\mu} & U_{\mu}(\mathbf{r}) \\
U^{\dagger}_{\mu}(\mathbf{r}) & -\hbar v_{F}(\hat{\mathbf{k}}-\mathbf{K}^{\mu}_{2})\cdot\mathbf{\sigma}_{\mu}
\end{pmatrix}\;,
\label{eq:Ham}
\end{equation}
where $v_F$ represents the Fermi velocity, $\hat{\mathbf{k}}=-i\nabla$, and $\mathbf{\sigma}_{\mu}=(\mu\sigma_{x},\sigma_y)$ denote the Pauli matrix in the sublattice space. Our Hamiltonian is expanded near the Dirac point $\mathbf{K}^{\mu}_{l}$ ($l=1,2$), with $\mu=\mp$ standing for the $K/K'$ valley. The $\mathbf{U}_{\mu}(\mathbf{r})$ matrix refers to the interlayer coupling which introduces a smooth moir\'e potential,
\begin{equation}
U_{\mu}(\mathbf{r})=\begin{pmatrix}
u_0 g_{\mu}(\mathbf{r}) & u_0'g_{\mu}(\mathbf{r}-\mu\mathbf{r}_{AB})\\
u_0'g_{\mu}(\mathbf{r}+\mu\mathbf{r}_{AB}) & u_0 g_{\mu}(\mathbf{r})
\end{pmatrix}e^{i\mu\Delta \mathbf{K}\cdot \mathbf{r}}\;,
\label{eq:u}
\end{equation}
where $\mathbf{r}_{AB}\!=\!(\sqrt{3}L_s/3,0)$,  $u_0^{\prime}$ and $u_0$ refer to the intersublattice and intrasublattice interlayer tunneling amplitudes, with $u_0'\!=\!0.0975\,$eV, and $u_0\!=\!0.0797$\,eV \cite{koshino-prx18}. $u_0$ is smaller than $u_0'$ due to the effects of atomic corrugations. $\Delta\mathbf{K}=\mathbf{K}-\mathbf{K}^{\prime}=(0,4\pi/3L_s)$ is the shift between the Dirac points of two layers. We define the phase factor $g(\mathbf{r})=\sum_{j=1}^{3}e^{-i\mu\mathbf{q}_j\cdot\mathbf{r}}$, with $\mathbf{q}_1=(0,4\pi/3L_s)$, $\mathbf{q}_2=(-2\pi/\sqrt{3}L_s,-2\pi/3L_s)$, and $\mathbf{q}_3=(2\pi/\sqrt{3}L_s,-2\pi/3L_s)$.

The moir\'e potential $U_M(\mathbf{r})$ preserves the moir\'e translational symmetry with real-space primitive lattice vectors $\mathbf{R}_1=(\sqrt{3}L_s/2,-L_s/2)$ and $\mathbf{R}_2=(\sqrt{3}L_s/2,L_s/2)$, and the corresponding reciprocal vectors are denoted as $\mathbf{g}_1=(2\pi/(\sqrt{3}L_s),-2\pi/L_s)$ and $\mathbf{g}_2=(2\pi/(\sqrt{3}L_s),2\pi/L_s)$. 
In order to study the density wave states, we consider the possibility that the system spontaneously breaks the moir\'e translational symmetry forming a doubled moir\'e supercell and a tripled $\sqrt{3}\times\sqrt{3}$ moir\'e supercell, where the lattice vectors are denoted by  red arrows (doubled cell) and blue arrows (tripled cell) in Fig.~1(a) of main text. For a generic case, we consider an enlarged moir\'e supercell characterized by $\mathbf{R}^{s}_{1}=n_{11}\mathbf{R}_1+n_{12}\mathbf{R}_2$,
and $\mathbf{R}^{s}_{2}=n_{21}\mathbf{R}_1+n_{22}\mathbf{R}_2$, where $n_{11}$, $n_{12}$, $n_{21}$, and $n_{22}$ are integers characterizing the enlarged moir\'e supercell. The reciprocal  vectors of the primitive  moir\'e cell are denoted by $\mathbf{b_1}$ and $\mathbf{b_2}$, then the reciprocal moir\'e vectors of the $M_s$-time  moir\'e supercell can be written as
\begin{equation}
\begin{split}
\mathbf{G_1}=(n_{22}\mathbf{b_1}-n_{21}\mathbf{b_2})/N_c,\\
\mathbf{G_2}=(n_{11}\mathbf{b_2}-n_{12}\mathbf{b_1})/N_c,
\end{split}
\label{eq:reciprocal}
\end{equation}
where $N_c=n_{11}n_{22}-n_{12}n_{21}$ (we take the convention that $N_c>0$) denotes the ratio between the area of the enlarged moir\'e supercell and the primitive moir\'e cell. To be specific, $n_{11}=n_{22}=1$ and $n_{12}=n_{21}=0$ refer to the primitive moir\'e cell of TBG. If $n_{11}=n_{22}=n_{12}=-n_{21}=1$, this group of indices represent the doubled  moir\'e supercell. The tripled $\sqrt{3}\times\sqrt{3}$ moir\'e supercell is characterized by $n_{11}=n_{12}=-n_{21}=1$ and $n_{22}=2$.  
The inverse of Eq.~(\ref{eq:reciprocal}) reads
\begin{equation}
\begin{split}
\mathbf{g_1}=n_{11}\mathbf{G_1}+n_{21}\mathbf{G_2},\\
\mathbf{g_2}=n_{12}\mathbf{G_1}+n_{22}\mathbf{G_2}.
\end{split}
\label{eq:reciprocal2}
\end{equation}
We have performed Hartree-Fock calculations for primitive, doubled, and tripled moir\'e supercells, using the expressions of the reciprocal vectors given in Eqs.~(\ref{eq:reciprocal})-(\ref{eq:reciprocal2}).  In our calculations, a 9$\times$9 grid in the reciprocal space is used for the primitive cell,  a 13$\times$7 grid is used for the doubled moir\'e cell, and a 11$\times$11 grid is used for the tripled moir\'e cell.  A 24$\times$24 $\mathbf{k}$-point mesh is adopted for the moir\'e Brillouin zone of the primitive cell, a 24$\times$14 $\mathbf{k}$-point mesh is adopted for the Brillouin zone of the doubled moir\'e cell, and a 12$\times$12 $\mathbf{k}$-point mesh is used for the Brillouin zone of the  tripled  moir\'e cell.

\vspace{12pt}
\begin{center}
\textbf{\large \II\ The Hartree-Fock method and remote-band Coulomb potentials}
\end{center}

The inter-site Coulomb interactions in the TBG system is expressed as,
\begin{equation}
H_C\!=\!\frac{1}{2N_s}\sum _{\alpha \alpha ^{\prime}}\sum _{\mathbf{k} \mathbf{k^{\prime}}\mathbf{q}}\sum _{\sigma \sigma ^{\prime}}\,V(\mathbf{q})\,\hat{c}^{\dagger}_{\mathbf{k+q},\alpha \sigma}\,\hat{c}^{\dagger}_{\mathbf{k^{\prime}-q}, \alpha ^{\prime}\sigma ^{\prime}}
\,\hat{c}_{\mathbf{k^{\prime}},\alpha ^{\prime}\sigma ^{\prime}}\,\hat{c}_{\mathbf{k},\alpha \sigma}
\label{eq:coulomb}
\end{equation}
where $\mathbf{k}$ and $\mathbf{q}$ represent atomic wavevectors defined in the Brillouin zone of graphene. The layer and sublattice indices are denoted by $\alpha$,  and $\sigma$ refers to the spin index. The $\hat{c}_{\mathbf{k},\alpha\sigma}$ and $\hat{c}_{\mathbf{k},\alpha\sigma}^{\dagger}$ refer to the electron annihilation and creation operators.  One can further expand the wavevectors around the Dirac points, which distinguish the Coulomb interaction into an intravalley part and an intervalley part, where the dominant intra-valley Coulomb interaction is expressed as, 
\begin{equation}
\begin{split}
H_{C}^{\rm{intra}}=\frac{1}{2N_s}\sum_{\alpha\alpha '}\sum_{\mu\mu ',\sigma\sigma '}\sum_{\mathbf{k}\mathbf{k} '\mathbf{q}}\,V(\mathbf{q})\,
\hat{c}^{\dagger}_{\mathbf{k}+\mathbf{q},\mu \sigma \alpha} \hat{c}^{\dagger}_{\mathbf{k}'-\mathbf{q},\mu '\sigma '\alpha '}
\hat{c}_{\mathbf{k}',\mu '\sigma '\alpha '}\hat{c}_{\mathbf{k},\mu \sigma \alpha}\;,
\label{eq:h-intra}
\end{split}
\end{equation}
where $\mu,\mu'=\pm$ denote the valley indices.
The double-gate screened Coulomb interaction
\begin{equation}
V(\mathbf{q})=\frac{e^2\tanh(|\mathbf{q}|d_s)}{\,2\Omega _M\epsilon_{\textrm{BN}} \epsilon _0 |\mathbf{q}|}
\label{eq:double-gate}
\end{equation}
in which $\Omega _M$ is the area of moir\'e supercell, $d_s$ is the screening length 400\AA{} and $\epsilon_{\textrm{BN}}=4$ refers to the dielectric constant of the hexagonal boron nitride substrates.

We further transform the original basis to the band basis
\begin{equation}
\hat{c}_{\mathbf{k},\mu\alpha\sigma}=\sum _n C_{\mu \alpha \mathbf{G},n\kt}\,\hat{c}_{\mu\sigma , n\widetilde{\mathbf{k}}}\;,
\label{eq:transform}
\end{equation}
in which $C_{\mu\alpha \mathbf{G},n\kt}$ is the wavefunction coefficient of the $n$th Bloch eigenstate at $\widetilde{\mathbf{k}}$ of valley $\mu$: $\vert\psi_{\mu,n\widetilde{\mathbf{k}}}\rangle=\sum_{\alpha\mathbf{G}}C_{\mu\alpha\mathbf{G},n\kt}\vert \kt+\G,\mu\alpha \rangle$, with $\mathbf{G}$ denoting the reciprocal vector. Under this transformation, the intra-valley Coulomb interaction can be written as
\begin{equation}
H^{\rm{intra}}
=\frac{1}{2N_s}\sum _{\widetilde{\mathbf{k}} \widetilde{\mathbf{k}}'\widetilde{\mathbf{q}}}\sum_{\substack{\mu\mu' \\ \sigma\sigma'}}\sum_{\substack{nm\\ n'm'}}\left(\sum _{\mathbf{Q}}\,V(\mathbf{Q}+\mathbf{\widetilde{q}})\,\Omega^{\mu \sigma,\mu'\sigma'}_{nm,n'm'}(\widetilde{\mathbf{k}},\widetilde{\mathbf{k}}',\widetilde{\mathbf{q}},\mathbf{Q})\right)\,\hat{c}^{\dagger}_{\mu\sigma,n\widetilde{\mathbf{k}}+\widetilde{\mathbf{q}}} \hat{c}^{\dagger}_{\mu'\sigma',n'\widetilde{\mathbf{k}}'-\widetilde{\mathbf{q}}}\,\hat{c}_{\mu'\sigma',m'\widetilde{\mathbf{k}}'}\,\hat{c}_{\mu\sigma,m\widetilde{\mathbf{k}}}
\label{eq:Hintra-band}
\end{equation}
%
in which the form factor $\Omega ^{\mu \sigma,\mu'\sigma'}_{nm,n'm'}$ is written as
\begin{equation}
\Omega ^{\mu \sigma,\mu'\sigma'}_{nm,n'm'}(\widetilde{\mathbf{k}},\widetilde{\mathbf{k}}',\widetilde{\mathbf{q}},\mathbf{Q})\,
=\sum _{\alpha\alpha'\mathbf{G}\mathbf{G}'}C^*_{\mu\sigma\alpha\mathbf{G}+\mathbf{Q},n\widetilde{\mathbf{k}}+\widetilde{\mathbf{q}}}C^*_{\mu'\sigma'\alpha'\mathbf{G}'-\mathbf{Q},n'\widetilde{\mathbf{k}}'-\widetilde{\mathbf{q}}}C_{\mu'\sigma'\alpha'\mathbf{G}',m'\widetilde{\mathbf{k}}'}C_{\mu\sigma\alpha\mathbf{G},m\widetilde{\mathbf{k}}}
\end{equation}
Here the wavevector $\mathbf{q}$ is decomposed as $\mathbf{q}=\mathbf{Q}+\mathbf{\widetilde{q}}$, $\mathbf{k}=\mathbf{G}+\mathbf{\widetilde{k}}$ where $\mathbf{G}$ or $\mathbf{Q}$ is a moir\'e reciprocal vector and $\mathbf{\widetilde{k}}$ or $\mathbf{\widetilde{q}}$ remarks the wavevector in the moir\'e Brillouin zone.  
When we perform extended Hartree-Fock calculations for the moir\'e superlattices with doubled or tripled primitive cells, $\mathbf{G}$ in the above equation refers to the  moir\'e reciprocal vectors of the enlarged supercell, $\widetilde{\mathbf{k}}$ or $\widetilde{\mathbf{q}}$ refers to a wavevector in the folded moir\'e Brillouin zone, as shown in Fig.~1(b) of main text.

Since intravalley Coulomb interactions are the leading ones in TBG, from now on we only consider the intravalley Coulomb interaction under band basis. We make Hartree-Fock approximation to Eq.~(\ref{eq:Hintra-band}) to decompose the two-particle interactions into a superposition of the Hartree and Fock mean-field single-particle Hamiltonians, where the Hartree term is expressed as
\begin{equation}
\begin{split}
H_H^{\rm{intra}}=&\frac{1}{2N_s}\sum _{\widetilde{\mathbf{k}} \widetilde{\mathbf{k}}'}\sum _{\substack{\mu\mu'\\ \sigma\sigma'}}\sum_{\substack{nm\\ n'm'}}\left(\sum _{\mathbf{Q}} V(\mathbf{Q})\Omega ^{\mu \sigma,\mu'\sigma'}_{nm,n'm'}(\widetilde{\mathbf{k}},\widetilde{\mathbf{k}}',0,\mathbf{Q})\right)\\
&\times \left(\langle \hat{c}^{\dagger}_{\mu\sigma,n\widetilde{\mathbf{k}}}\hat{c}_{\mu\sigma,m\widetilde{\mathbf{k}}}\rangle \hat{c}^{\dagger}_{\mu'\sigma',n'\widetilde{\mathbf{k}}'}\hat{c}_{\mu'\sigma',m'\widetilde{\mathbf{k}}'} + \langle \hat{c}^{\dagger}_{\mu'\sigma',n'\widetilde{\mathbf{k}}'}\hat{c}_{\mu'\sigma',m'\widetilde{\mathbf{k}}'}\rangle \hat{c}^{\dagger}_{\mu\sigma,n\widetilde{\mathbf{k}}}\hat{c}_{\mu\sigma,m\widetilde{\mathbf{k}}}\right)
\end{split}
\label{eq:hartree}
\end{equation}
%
and the Fock term is
%
\begin{equation}
\begin{split}
H_F^{\rm{intra}}=&-\frac{1}{2N_s}\sum _{\widetilde{\mathbf{k}} \widetilde{\mathbf{k}}'}\sum _{\substack{\mu\mu'\\ \sigma\sigma'}}\sum_{\substack{nm\\ n'm'}}\left(\sum _{\mathbf{Q}} V(\widetilde{\mathbf{k}}’-\widetilde{\mathbf{k}}+\mathbf{Q})\Omega ^{\mu \sigma,\mu'\sigma'}_{nm,n'm'}(\widetilde{\mathbf{k}},\widetilde{\mathbf{k}}',\widetilde{\mathbf{k}}’-\widetilde{\mathbf{k}},\mathbf{Q})\right)\\
&\times \left(\langle \hat{c}^{\dagger}_{\mu\sigma,n\widetilde{\mathbf{k}}'}\hat{c}_{\mu'\sigma',m'\widetilde{\mathbf{k}}'}\rangle \hat{c}^{\dagger}_{\mu'\sigma',n'\widetilde{\mathbf{k}}}\hat{c}_{\mu\sigma,m\widetilde{\mathbf{k}}} + \langle \hat{c}^{\dagger}_{\mu'\sigma',n'\widetilde{\mathbf{k}}}\hat{c}_{\mu\sigma,m\widetilde{\mathbf{k}}}\rangle \hat{c}^{\dagger}_{\mu\sigma,n\widetilde{\mathbf{k}}'}\hat{c}_{\mu'\sigma',m'\widetilde{\mathbf{k}}'}\right).
\end{split}
\label{eq:fock}
\end{equation}

We further project the interaction Hamiltonian (with Hartree-Fock approximations) onto the flat-band subspace. However, the treatment of the remote bands is very tricky when making such a projection. Since the remote bands below the charge neutrality point (CNP) are all occupied, which can interact with the electrons occupying the flat bands, and such interactions are pointed out to be important in determining the ground states and low-energy excitations in magic-angle TBG \cite{Bernevig-tbg3-arxiv20,Lian-tbg4-arxiv20,kang-cascade-tbg-arxiv21}. In particular, in Ref.~\onlinecite{Bernevig-tbg3-arxiv20}, it is proposed that one can conveniently write the interaction as 
\begin{equation}
H_{C}^{\prime}=\frac{N_s}{2}\sum_{\mathbf{q}}V(\mathbf{q})\delta \hat{\rho}(\mathbf{q})\delta \hat{\rho}(-\mathbf{q}),
\label{eq:HC}
\end{equation}
where $\delta \hat{\rho}(\mathbf{q})=\delta \hat{\rho}(\widetilde{\mathbf{q}}+\mathbf{Q})$ is defined as
\begin{equation}
\delta\rho(\mathbf{q})=\frac{1}{N_s}\sum_{\widetilde{\mathbf{k}}}\sum_{\mu\sigma\alpha\mathbf{G}}\Big(\,c^{\dagger}_{\mu\sigma\alpha\mathbf{G}+\mathbf{Q}+\widetilde{\mathbf{k}}+\widetilde{\mathbf{q}}}c_{\mu\sigma\alpha,\mathbf{G}+\widetilde{\mathbf{k}}}-(1/2) \delta_{\widetilde{\mathbf{q}},\mathbf{0}}\delta_{\widetilde{\mathbf{Q}},\mathbf{0}}\,\Big).    
\end{equation}
Again, the $\mu$, $\sigma$, and $\alpha$ indices refer to the valley, spin, and layer/sublattice indices respectively. $\mathbf{G}$ or $\mathbf{Q}$ denotes to a moir\'e reciprocal vector and $\widetilde{\mathbf{k}}$ or $\widetilde{\mathbf{q}}$ denotes a wavevector within the moir\'e Brillouin zone. In the case of doubled or tripled moir\'e supercell,$\mathbf{G}$ and $\mathbf{Q}$ would correspond to the  reciprocal vectors of the enlarged moir\'e  supercells, and $\widetilde{\mathbf{k}}$ and $\widetilde{\mathbf{q}}$ are wavevectors within the folded moir\'e Brillouin zone as shown in Fig.~1(b) of main text.

Note that if the full Hilbert space is included, then Eq.~(\ref{eq:HC}) is equivalent to the normal-ordered interaction (Eq.~(\ref{eq:h-intra})) up to a constant chemical potential term. However, 
if the interaction is projected onto the flat bands, the Hamiltonian in Eq.~(\ref{eq:HC}) becomes different from the normal-ordered one. 
It turns out that if one makes Hartree-Fock approximations, then the remote-band Hartree-Fock potential acted on the flat bands is exactly the difference between Eq.~(\ref{eq:HC}) and the normal-ordered one by virtue of the $C_{2z}\mathcal{T}$ symmetry and a particle-hole symmetry of the continuum model \cite{Bernevig-tbg3-arxiv20}.
To be specific, the remote-band  Hartree-Fock potential can be expressed in band basis as:
\begin{equation}
\begin{split}
\Delta H_{I}=&-\frac{1}{2N_s}\sum_{\mathbf{k}\mu s}\sum_{nm} \left(\sum_{\mathbf{Q}}V(\mathbf{Q})\,\sum_{\substack{\widetilde{\mathbf{k}}'\mu' \\n's'}}\Omega^{\mu,\mu'}_{nm,n'n'}(\widetilde{\mathbf{k}},\widetilde{\mathbf{k}}',0,\mathbf{Q})\right)\hat{c}^{\dagger}_{\mu s,n\widetilde{\mathbf{k}}}\hat{c}_{\mu s,m\widetilde{\mathbf{k}}}\\
 &+\frac{1}{2N_s}\sum_{\mathbf{k}\mu s}\sum_{nm} \left(\sum_{\mathbf{Q}\widetilde{\mathbf{q}}}V(\mathbf{Q}+\widetilde{\mathbf{q}})\,\sum_{n'}\Omega^{\mu,\mu}_{nn',n'm}(\widetilde{\mathbf{k}}-\widetilde{\mathbf{q}},\widetilde{\mathbf{k}},\widetilde{\mathbf{q}},\mathbf{Q})\right)\hat{c}^{\dagger}_{\mu s,n\widetilde{\mathbf{k}}}\hat{c}_{\mu s,m\widetilde{\mathbf{k}}}\;,
\end{split}
\label{eq:remote}
\end{equation}
where the summations of the band indices $n, m, n',$ and $m'$ are restricted to the flat-band subspace. In our calculations, the Coulomb interactions are projected onto the two flat bands per spin per valley, and the remote-band Hartree-Fock potentials given in Eq.~(\ref{eq:remote}) have been included in our calculations.

\vspace{12pt}
\begin{center}
\textbf{\large \III\ The screening effects and constraint random phase approximation}
\end{center}

In the previous definition of Coulomb interactions (Eq.~(\ref{eq:h-intra})), a double-gate screened interaction has been considered as shown in Eq.~(\ref{eq:double-gate}), in which only the screening effects from the top and bottom metallic gates have been included. The gate-screened Coulomb interactions are further projected onto the flat bands of the TBG system. In reality, the Coulomb interaction between two electrons occupying the flat bands can also be screened due to the virtual excitations of particle-hole pairs from the remote bands. We characterize such screening effects using the constrained random phase approximation (cRPA). In particular, in the flat-band subspace, the  Coulomb interaction including such screening effects is expressed as
\begin{align}
H^{\textrm{cRPA}}_{int}= \frac{1}{2N_{s}}\sum_{\widetilde{\mathbf{k}}\widetilde{\mathbf{k}^{\prime}}\widetilde{\mathbf{q}}}\sum_{\substack{\mu,\mu^{\prime} \\ \sigma,\sigma^{\prime}}}\sum_{\substack{n n^{\prime} \\ m m^{\prime} }} V^{\textrm{cRPA}}(\widetilde{\mathbf{q}})_{\widetilde{\mathbf{k}}\mu n m,\widetilde{\mathbf{k}^{\prime}}\mu^{\prime}n^{\prime}m^{\prime}}\hat{c}^{\dagger}_{\mu\sigma,n \widetilde{\mathbf{k}}+\widetilde{\mathbf{q}}} \hat{c}^{\dagger}_{\mu^{\prime}\sigma^{\prime},n^{\prime} \widetilde{\mathbf{k}^{\prime}}} \hat{c}_{\mu^{\prime}\sigma^{\prime},m^{\prime} \widetilde{\mathbf{k}^{\prime}}+\widetilde{\mathbf{q}}}\hat{c}_{\mu\sigma,m \widetilde{\mathbf{k}}}
\end{align}
in which the cRPA screened interaction in the flat-band basis $V^{\textrm{cRPA}}(\widetilde{\mathbf{q}})_{\widetilde{\mathbf{k}}\mu n m,\widetilde{\mathbf{k}^{\prime}}\mu^{\prime}n^{\prime}m^{\prime}}$ is expressed as
\begin{align}
V^{\textrm{cRPA}}(\widetilde{\mathbf{q}})_{\widetilde{\mathbf{k}}\mu n m,\widetilde{\mathbf{k}^{\prime}}\mu^{\prime}n^{\prime}m^{\prime}}=\sum_{\mathbf{Q},\mathbf{Q}^{\prime}}\lambda_{\widetilde{\mathbf{k}}\mu n m,\mathbf{Q}}(\widetilde{\mathbf{q}})V(\widetilde{\mathbf{q}})_{\mathbf{Q},\mathbf{Q}^{\prime}}\lambda^{\dagger}_{\mathbf{Q}^{\prime},\widetilde{\mathbf{k}^{\prime}}\mu^{\prime}m^{\prime}n^{\prime}}\;,
\end{align}
where $\lambda_{\widetilde{\mathbf{k}}\mu n m,\mathbf{Q}}$ is defined as
\begin{align}
\lambda_{\widetilde{\mathbf{k}}\mu n m,\mathbf{Q}}(\widetilde{\mathbf{q}})=\sum_{\alpha \mathbf{G}}C^{*}_{\mu\alpha \mathbf{G}+\mathbf{Q},n}(\widetilde{\mathbf{k}}+\widetilde{\mathbf{q}})C_{\mu\alpha \mathbf{G},m}(\widetilde{\mathbf{k}})
\end{align}

The cRPA screened Coulomb interaction in the band basis $V^{\textrm{cRPA}}(\widetilde{\mathbf{q}})_{\mathbf{Q},\mathbf{Q}^{\prime}}$ is expressed as 
\begin{equation}
\begin{split}
&V^{\textrm{cRPA}}_{\widetilde{\mathbf{k}}\mu n m,\widetilde{\mathbf{k}^{\prime}}\mu^{\prime} m^{\prime} n^{\prime}}(\widetilde{\mathbf{q}})= V_{\widetilde{\mathbf{k}}\mu n m,\widetilde{\mathbf{k}^{\prime}}\mu^{\prime} m^{\prime} n^{\prime}}(\widetilde{\mathbf{q}})-\frac{2}{N_{s}}\sum_{n_{1}m_{1}}'\sum_{\widetilde{\mathbf{k}}_1\mu_{1}}{V_{\widetilde{\mathbf{k}}\mu n m,\widetilde{\mathbf{k}}_{1}\mu_{1} m_{1} n_{1}}(\widetilde{\mathbf{q}})\chi^{0}_{\widetilde{\mathbf{k}}_{1}\mu_{1}m_{1}n_{1}}(\widetilde{\mathbf{q}})V_{\widetilde{\mathbf{k}}_{1}\mu_{1} m_{1} n_{1},\widetilde{\mathbf{k}^{\prime}}\mu^{\prime} n^{\prime} m^{\prime}}}(\widetilde{\mathbf{q}})\\
&+ \left(\frac{-2}{N_s}\right)^{2}\sum_{\substack{n_{1}m_{1}\\ n_{2}m_{2}}}'\sum_{\mu_{1}\mu_{2}}\sum_{\widetilde{\mathbf{k}}_{1}\widetilde{\mathbf{k}_{2}}}V_{\widetilde{\mathbf{k}}\mu n m,\widetilde{\mathbf{k}}_{1}\mu_{1} m_{1} n_{1}}(\widetilde{\mathbf{q}})\chi^{0}_{\widetilde{\mathbf{k}}_{1}\mu_{1}m_{1}n_{1}}(\widetilde{\mathbf{q}})V_{\widetilde{\mathbf{k}}_{1}\mu_{1} m_{1} n_{1},\widetilde{\mathbf{k}}_{2}\mu_{2} m_{2} n_{2}}(\widetilde{\mathbf{q}})\chi^{0}_{\widetilde{\mathbf{k}}_{2}\mu_{2}m_{2}n_{2}}(\widetilde{\mathbf{q}})V_{\widetilde{\mathbf{k}}_{2}\mu_{2} m_{2} n_{2},\widetilde{\mathbf{k}^{\prime}}\mu^{\prime} n^{\prime} m^{\prime}}(\widetilde{\mathbf{q}})+ \dots\;,
\end{split}
\end{equation}
where $V_{\widetilde{\mathbf{k}}\mu n m,\widetilde{\mathbf{k}^{\prime}}\mu^{\prime} m^{\prime} n^{\prime}}(\widetilde{\mathbf{q}})$ is the double-gate screened Coulomb interaction projected onto the flat bands, i.e., $V_{\widetilde{\mathbf{k}}\mu n m,\widetilde{\mathbf{k}^{\prime}}\mu^{\prime} m^{\prime} n^{\prime}}(\widetilde{\mathbf{q}})=\sum_{\mathbf{Q}}\,V(\mathbf{Q}+\widetilde{q})\,\Omega_{nm,n'm'}^{\mu,\mu'}(\widetilde{\mathbf{k}},\widetilde{\mathbf{k}}',\widetilde{\mathbf{q}},\mathbf{Q})$, as shown in Eq.~(\ref{eq:Hintra-band}). The band indices $m,m',n,n'$ refer to the flat-band indices. However,
the summation over band indices $\sum_{n_1 m_1}'$ with a superscript $'$ means that the band summation is restricted: $n_1$ and $m_1$ cannot be both the flat band indices. In other words, we consider the screening effects to the flat-band electrons from the virtual excitations of particle-hole pairs  through the following three processes: the particle-hole pairs are excited  (\i) from the remote bands below CNP to those above CNP,  (\ii) from remote bands below CNP to the flat bands, and (\iii) from the flat bands to remote bands above CNP. We exclude the particle-hole excitations within the flat-band subspace.  This is called the constrained random phase approximation \cite{pollet-tbg-crpa-prb20,wehling-tbg-prb19}.

The static bare susceptibility in the band basis is expressed as
\begin{align}
\chi^{0}_{\widetilde{\mathbf{k}}\,\mu\,m\,n}(\widetilde{\mathbf{q}})=\frac{f(E_{\mu,m\widetilde{\mathbf{k}}+\widetilde{\mathbf{q}}})-f(E_{\mu,n\widetilde{\mathbf{k}}})}{E_{\mu,n\widetilde{\mathbf{k}}}-E_{\mu,m\widetilde{\mathbf{k}}+\widetilde{\mathbf{q}}}}
\end{align}
One can also define the bare susceptibility in the basis of the transferred reciprocal vectors 
\begin{align}
\chi^{0}(\widetilde{\mathbf{q}})_{\mathbf{Q}_{1},\mathbf{Q}_{2}}=\frac{2}{N_{s}}\sum_{\widetilde{\mathbf{k}}_1}\sum _{\mu_{1}m_{1}n_{1}}'\lambda^{\dagger}(\widetilde{\mathbf{q}})_{\mathbf{Q}_{1},\widetilde{\mathbf{k}}_{1}\mu_{1}m_{1}n_{1}}\,\chi^{0}_{\widetilde{\mathbf{k}}_{1}\mu_{1}m_{1}n_{1}}(\widetilde{\mathbf{q}})\,\lambda(\widetilde{\mathbf{q}})_{\widetilde{k_{1}}\mu_{1}m_{1}n_{1},\mathbf{Q}_{2}}\;.
\label{chi0}
\end{align}
It should be noted that the summation over $\widetilde{\mathbf{k}}$ can be difficult to converge when there is any van Hove singularity in the energy bands, therefore we have adopted the analytic linear interpolation method \cite{interpolation-prb75} to perform the summation of $\widetilde{\mathbf{k}}$ points. 
Then the cRPA screened Coulomb interaction can be re-written as:
\begin{equation}
V^{\textrm{cRPA}}_{\widetilde{\mathbf{k}}\mu n m,\widetilde{\mathbf{k}^{\prime}}\mu^{\prime} m^{\prime} n^{\prime}}(\widetilde{\mathbf{q}})=\sum_{\mathbf{Q},\mathbf{Q}^{\prime}}\lambda_{\widetilde{\mathbf{k}}\mu n m,\mathbf{Q}}(\widetilde{\mathbf{q}})\left[\hat{V}(\widetilde{\mathbf{q}})\cdot\left(1+\chi^{0}(\widetilde{\mathbf{q}})\cdot \hat{V}(\widetilde{\mathbf{q}})\right)^{-1}\right]_{\mathbf{Q},\mathbf{Q}^{\prime}}\lambda^{\dagger}_{\mathbf{Q}^{\prime},\widetilde{\mathbf{k}^{\prime}}\mu^{\prime}m^{\prime}n^{\prime}}(\widetilde{\mathbf{q}})\;.
\label{eq:Vcrpa}
\end{equation}
In the above equation, the double-gate screened Coulomb interaction $V(\mathbf{q})=V(\widetilde{\mathbf{q}}+\mathbf{Q})$ (Eq.~(\ref{eq:double-gate})) is written in matrix form: $\hat{V}(\widetilde{\mathbf{q}})_{\mathbf{Q},\mathbf{Q}'}=V(\widetilde{\mathbf{q}}+\mathbf{Q})\delta_{\mathbf{Q},\mathbf{Q}'}$, and we define the dielectric matrix
\begin{equation}
\hat{\epsilon}_{\mathbf{Q},\mathbf{Q}^{\prime}}(\widetilde{\mathbf{q}})=\left(1+\chi^{0}(\widetilde{\mathbf{q}})\cdot \hat{V}(\widetilde{\mathbf{q}})\right)_{\mathbf{Q},\mathbf{Q}^{\prime}}
\label{eq:epsilon}
\end{equation}
The dielectric function $\epsilon(\widetilde{\mathbf{q}}+\mathbf{Q})$ is defined as the dominant diagonal element of Eq.~(\ref{eq:epsilon}),  and $\epsilon(\widetilde{\mathbf{q}}+\mathbf{Q})\times\epsilon_{\textrm{BN}}$ ($\epsilon_{\textrm{BN}}=4$) is plotted in Fig.~1(e) of main text.
Then, the cRPA screened Coulomb interaction is expressed as
\begin{align}
\bar{V}^{\textrm{cRPA}}(\widetilde{\mathbf{q}})=\hat{V}(\widetilde{\mathbf{q}})\cdot\left(\hat{\epsilon}(\widetilde{\mathbf{q}})\right)^{-1}
\end{align}
We  further project the cRPA screened Coulomb interactions onto the flat bands, as expressed in Eq.~(\ref{eq:Vcrpa}). 
After making Hartree-Fock approximations, the interaction Hamiltonian is decomposed into Hartree and Fock terms as shown in Eqs.~(\ref{eq:hartree})-(\ref{eq:fock}). We take $V^{\textrm{cRPA}}_{\widetilde{\mathbf{k}}\mu n m,\widetilde{\mathbf{k}^{\prime}}\mu^{\prime} m^{\prime} n^{\prime}}(\widetilde{\mathbf{q}}=\mathbf{0})$ in the Hartree term, and we let $V(\mathbf{k}'-\mathbf{k}+\mathbf{Q})\to V(\mathbf{k}'-\mathbf{k}+\mathbf{Q})/\epsilon(\mathbf{k}'-\mathbf{k}+\mathbf{Q})$ in the Fock term, where $V(\mathbf{q})$ is the double-gate screened Coulomb interaction (without the remote-band screening effects) given in Eq.~(\ref{eq:double-gate}).


\vspace{12pt}
\begin{center}
\textbf{\large \IV\ More Hartree-Fock results at integer fillings with preserved moir\'e translational symmetry}
\end{center}

\begin{figure}[!htbp]
\includegraphics[width=0.8\textwidth]{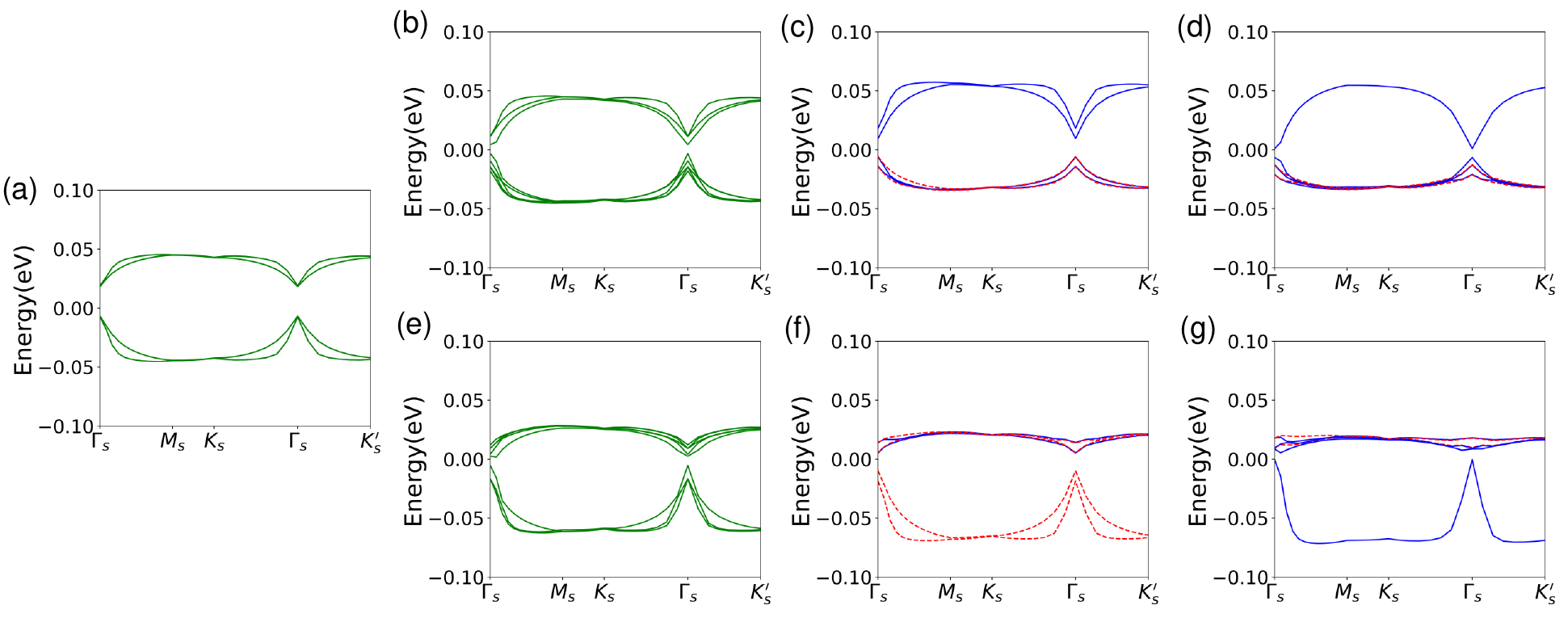}
\caption{~\label{figs01} The Hartree-Fock energy bands of twisted bilayer graphene with $\theta=1.05^{\circ}$, at the fillings  (a) $\nu$\,=\,0, (b) $\nu$=1, and (c) $\nu$\,=\,2, (d) $\nu$\,=\,3, (e) $\nu$\,=\,-1, (f) $\nu$\,=\,-2, and (g) $\nu\,=\,$-3, where the screening effects from the remote bands  and the remote-band Hartree-Fock potentials acted on the flat bands are involved. 
At the filling $\nu$\,=\,$\pm 2$ and $\nu$\,=\,$\pm 3$, the ground states are spin-valley polarized, thus the energy bands from two valley are marked with blue solid lines and red dash lines. For $\nu\!=\!0, \pm 1$, the ground states are intervalley coherent, and the energy bands are marked by green lines.}
\end{figure}

\begin{figure}[!htbp]
\includegraphics[width=0.8\textwidth]{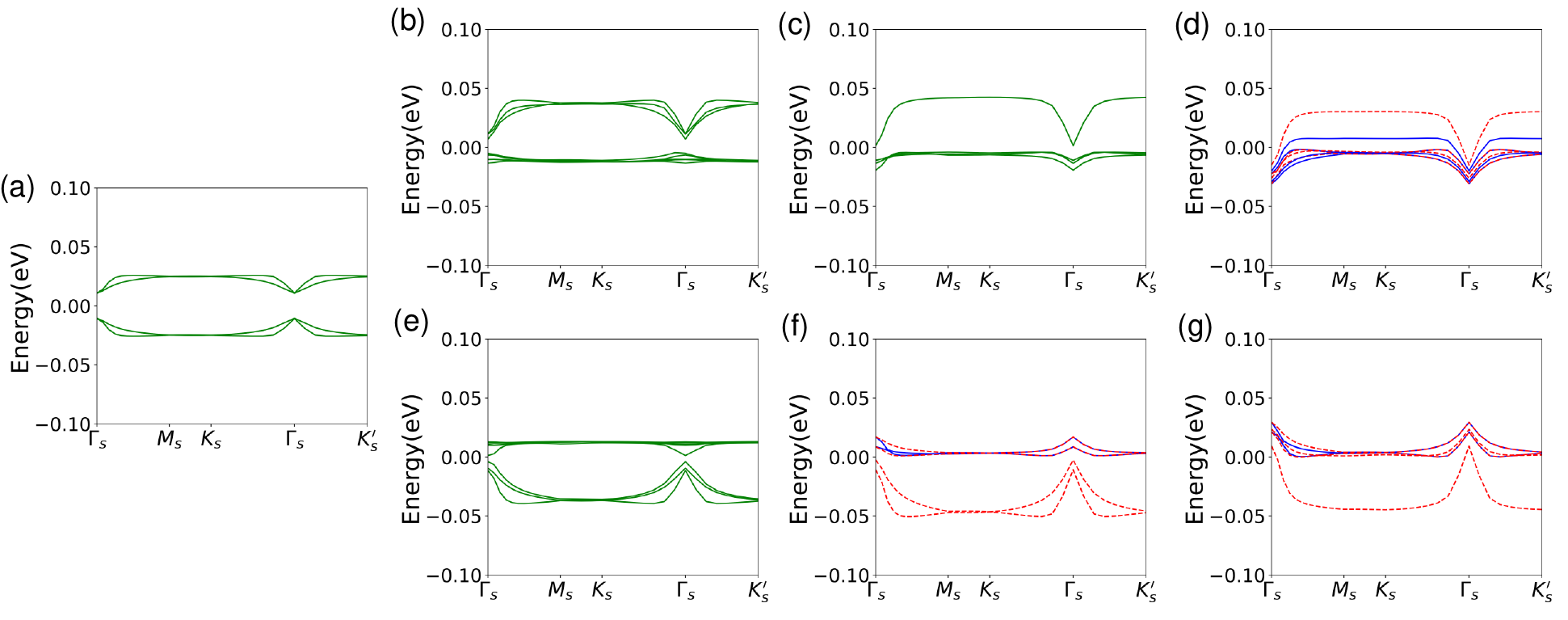}
\caption{~\label{figs02} The Hartree-Fock energy bands of twisted bilayer graphene with $\theta=1.05^{\circ}$, at the fillings  (a) $\nu$\,=\,0, (b) $\nu$=1, and (c) $\nu$\,=\,2, (d) $\nu$\,=\,3, (e) $\nu$\,=\,-1, (f) $\nu$\,=\,-2, and (g) $\nu\,=\,$-3, where only the remote-band Hartree-Fock potentials acted on the flat bands are involved and cRPA screening is ignored here.}
\end{figure}

As discussed in main text, at $\nu=0$, our calculations reveal that the ground state is a Kramers intervalley coherent (K-IVC) state characterized by order parameters $(\tau_x\sigma_y,\tau_y\sigma_y)$ \cite{zaletel-tbg-hf-prx20}, with  mixture of small valley polarization component $\langle\tau_z\rangle\approx 0.1$. 
In the chiral limit ($u_0=0$) with finite kinetic energy, a pure K-IVC state can be obtained at the CNP, which is to be discussed below. 
The correlated state at the $\nu$\,=\,$\pm 1$ is a Kramers intervalley coherent (K-IVC) state mixed with some valley polarization $\langle \tau_z \rangle=\!-0.85 (1.14)$ for $\nu\!=\!1 (-1)$. Such a state at $\nu=\pm 1$ can be intuitively interpreted as follows: two occupied Chern bands from the two valleys with opposite Chern numbers $\pm 1$ are coupled and trivialized by the K-IVC order, leaving one Chern band being occupied, thus the total Chern number is $\pm 1$.
The states at the $\nu$\,=\,$\pm 2$ and $\nu$\,=\,$\pm 3$  become fully spin-valley-polarized, with Chern number 0 and $\pm 1$ respectively.

The single-particle excitation spectra calculated by Hartree-Fock+cRPA method with the filling factors fixed at $\nu=-3, -2,-1 ,0 ,1, 2, 3$ (preserving moir\'e translational symmetry) are shown in Fig.~\ref{figs01}. For comparison, we also present the single-particle excitations from Hartree-Fock calculations without the cRPA screening effects at all the integer fillings  in Fig.~\ref{figs02}. We note that, including the cRPA screening effects, the ground states are gapped at all integer fillings, and the calculated energy gaps are shown in Table.~\ref{tables0}. On the other hand, without the cRPA screening effects, the Hartree-Fock ground states are gapped at fillings $\nu\!=\!0,\pm1,\pm2$, but becomes gapless with slight negative indirect gaps for $\nu\!=\!\pm3$ as shown in Fig.~\ref{figs02}. Moreover, we also find that the ground state at $\nu\!=\!0$ is a pure K-IVC state, thus we conclude that including the remote-band screening effects would introduce small valley polarization components into the K-IVC state at CNP.

\begin{table}[!t]
 \caption{Energy gaps for magic-angle TBG calculated by Hartree-Fock+cRPA method.\label{tables0}}
\begin{tabular}{c|c|c|c|c|c|c|c}
\hline
filling & $\nu\!=\!3$ & $\nu\!=\!2$ & $\nu\!=\!1$ & $\nu\!=\!0$ & $\nu\!=\!-1$ & $\nu\!=\!-2$ & $\nu\!=\!-3$ \\
\hline
gap (meV) & 7.1 & 15.3 & 7.5 & 25.0 & 6.1 & 14.1 & 4.4 \\
\hline
\end{tabular}
\end{table}

Moreover, we find that the single-particle spectrum calculated by the Hartree-Fock+cRPA method roughly preserves particle-hole symmetry  for $\nu\!=\!0$ (Fig.~\ref{figs01}(a)), 
but for $\nu\!=\!2$ (Fig.~\ref{figs01}(c)) and $\nu\!=\!3$ (Fig.~\ref{figs01}(d)), the  conduction bands are  much more dispersive than the valence bands. The energy spectra at $\nu=-1, -2, -3$, as shown in Fig.~\ref{figs01}(e)-(g), are exactly the opposite, with the valence bands being more dispersive than the conduction bands.  In a recent theoretical study \cite{kang-cascade-tbg-arxiv21}, Kang \textit{et al.} argued that such asymmetric single-particle dispersions in TBG are the origin of the cascade transitions observed in experiments \cite{yazdani-cascade-tbg,ilani-cascade-tbg}.

We introduce a parameter $0\leq\lambda\leq1$ to characterize the bandwidth of flat bands in the continuum model. When $\lambda = 0$, the kinetic energy is zero, which is called the flat-band limit. In the following, we will discuss about the ground states under different $u_0$ and $\lambda$ parameters at the CNP and $\nu\!=\!2$ filling.

\begin{figure}[!htbp]
\includegraphics[width=0.35\textwidth]{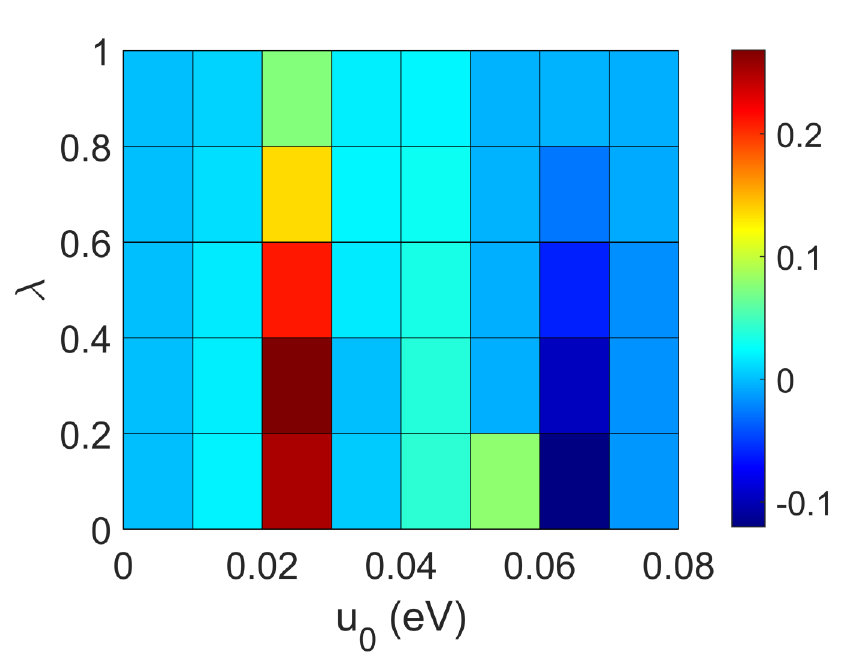}
\caption{~\label{CNP} The normalized valley polarization component $\langle \tau_z \rangle / \langle \tau_z \rangle _{\rm{VP}}$ in the K-IVC state at the CNP in which $\langle \tau_z \rangle _{\rm{VP}}$ is the valley polarization of pure valley-polarized state at the same parameter point. $\lambda$ represents the bandwidth of flat bands, and $\lambda\!=\!1$  corresponds to the realistic bandwidth. In the chiral limit, the calculated K-IVC states are all pure K-IVC phases without valley polarization.}
\end{figure}

At the CNP, in the chiral limit ($u_0\!=\!0$) and flat-band limit ($\lambda\!=\!0$), the ground state is highly degenerate due to the $U(4)\times U(4)$ symmetry of the interaction Hamiltonian \cite{zaletel-tbg-hf-prx20,Bernevig-tbg3-arxiv20}, which can also be obtained from our Hartree-Fock+cRPA calculations. 
If we stay in the flat-band limit ($\lambda\!=\!0$) and increase $u_0$ to break  chiral symmetry, the ground state consists of  a set of degenerate states of pure valley polarized (VP) state and K-IVC state,  by virtue of the $U(4)$ symmetry (including spin degrees of freedom) in the case of non-chiral flat limit \cite{Bernevig-tbg3-arxiv20}. Then in the non-chiral case, further increasing $\lambda$ (bandwidth) will raise up the energy of the VP state, and make the K-IVC (with slight mixtures of valley polarization) being the unique ground state as shown in Fig.~\ref{CNP}. In summary,   finite bandwidth of flat bands will lift up the energy of a pure VP state, and the breaking of chiral symmetry would induce a mild mixture of the VP order into the K-IVC state.

\begin{figure}[!htbp]
\includegraphics[width=0.6\textwidth]{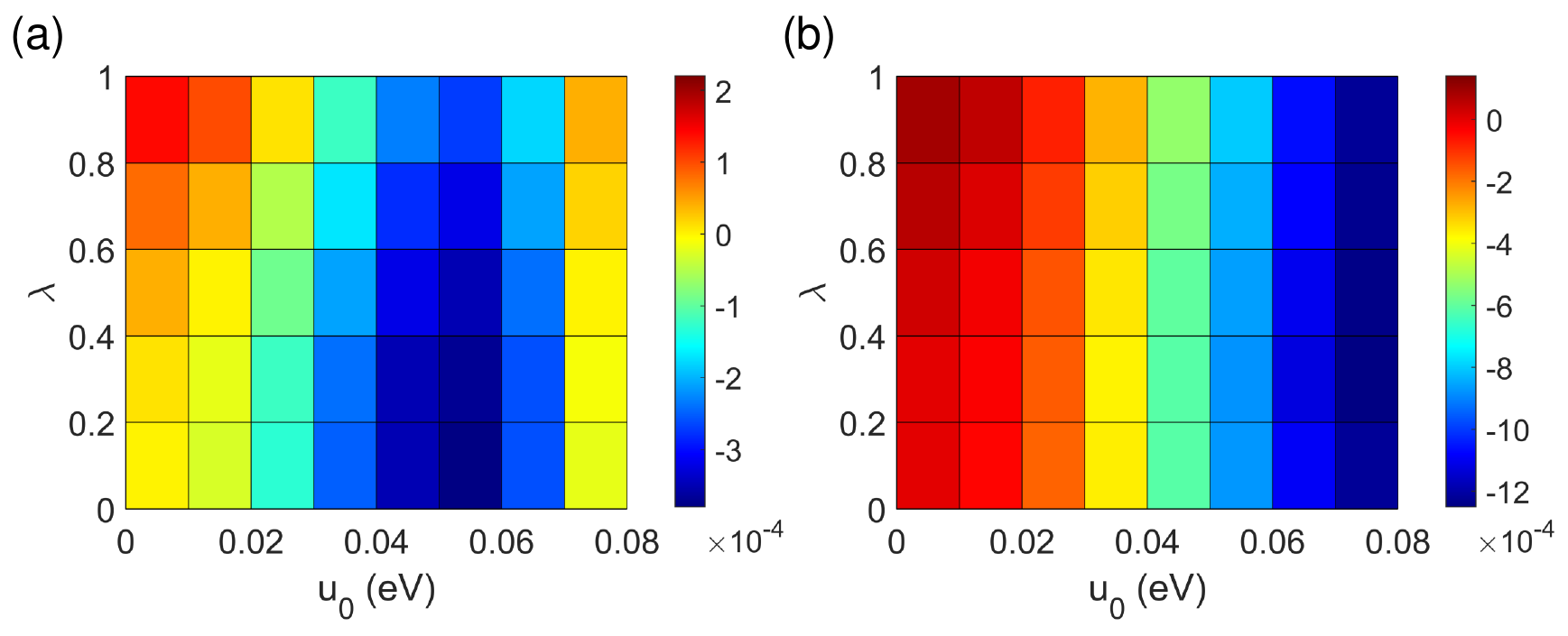}
\caption{~\label{filling2} The energy different between SVP phase and IVC phase $\Delta E=E_{SVP}-E_{IVC}$ (in units of eV): (a) with constant screening $\epsilon\!=\!10$,  and (b) with cRPA method.}
\end{figure}

As for the ground state at the $\nu\!=\!2$, we show the energy difference between spin-valley polarized (SVP) phase and intervalley coherent (IVC) phase $\Delta E=E_{SVP}-E_{IVC}$ with both constant screening $\epsilon\!=\!10$ and with the cRPA method  in Fig.~\ref{filling2}(a) and (b).  Here the IVC phase can involve both the K-IVC and time-reversal invariant IVC (T-IVC) phase, which are characterized by $(\tau_x\sigma_y,\tau_y\sigma_y)$ and $(\tau_x\sigma_x,\tau_y\sigma_x)$ respectively. 
With a fixed dielectric constant $\epsilon=10$, the HF ground states at $\nu\!=\!2$  with realistic parameters ($u_0\!=\!0.0797$\,eV and $\lambda\!=\!1$)  invovle three nearly degenerate states: the IVC state mixed with slight valley and sublattice polarizations ($\tau_i\sigma_j, \tau_z, \sigma_z$,$i=x,y$), a valley-sublattice polarized state (with order parameters $\tau_z, \sigma_z, \tau_z\sigma_z$), and a spin-sublattice polarized state (with order parameters $s_z,\tau_z\sigma_z,\tau_zs_z\sigma_z$). The energy difference between these states are no more than 10\,$\mu$eV.

\vspace{12pt}
\begin{center}
\textbf{\large \V\ More results about the Hartree-Fock calculations for the density wave states}
\end{center}

\begin{center}
\textbf{\large \V\ A Filling $\nu=7/2$}
\end{center}

\begin{figure}[!htbp]
\includegraphics[width=0.9\textwidth]{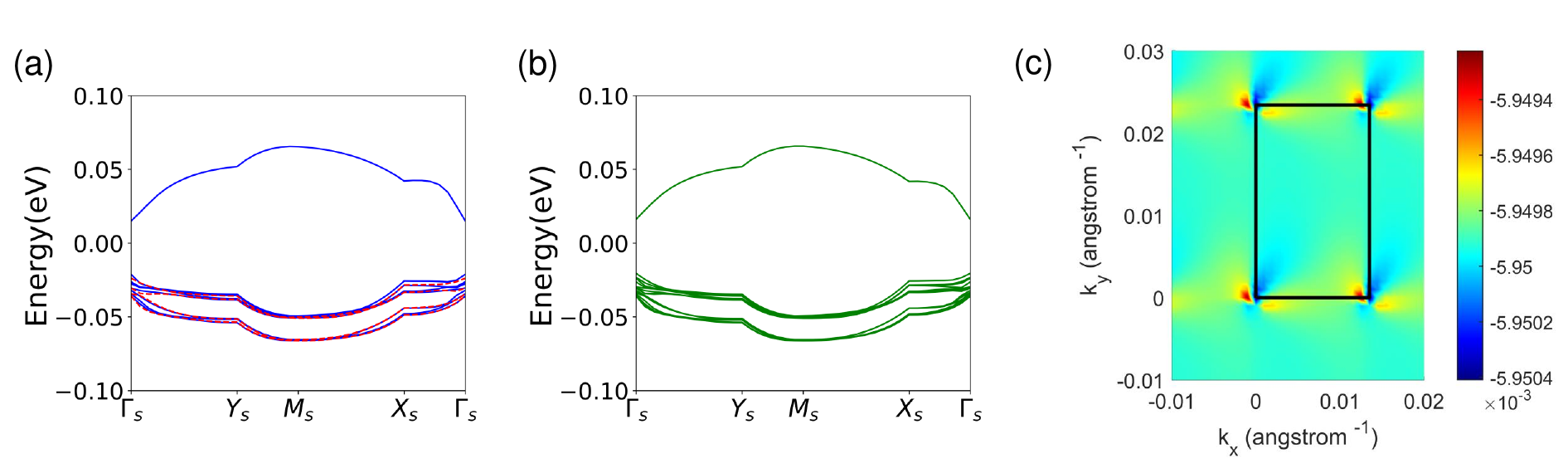}
\caption{~\label{figs04} The Hartree-Fock energy bands of 1.05$^{\circ}$--twisted bilayer graphene at the fillings $\nu$\,=\,7/2 (a) (SVP phase), $\nu$\,=\,7/2 (b) (K-IVC phase). The green lines represent the K-IVC state. In the spin-valley-polarized phase, the energy bands from two valley are remarked with blue solid lines and red dash lines. In (c) we also show the distribution of valley polarization in the reciprocal space $\langle \tau_z (\widetilde{\mathbf{k}}) \rangle$ at  filling $\nu\!=\!3$, where the rectangle marks the moir\'e Brillouin zone of the doubled moir\'e supercell.}
\end{figure}

The ground state at filling $\nu\!=\!7/2$ with doubled moir\'e supercell consists of  two nearly degenerate states: the K-IVC and the SVP state. Both of these two density-wave states with doubled  moir\'e supercell  break the $C_{3z}$ rotational symmetry.  The Hartree-Fock band structures of these two states at $\nu=7/2$ are shown in Fig.~\ref{figs04}(a) (for SVP state) and (b) (for K-IVC state).  The SVP state at 7/2 filling has a Chern number of 1, while the K-IVC state has a Chern number of 0. Thus, the K-IVC state at 7/2 filling is consistent with the experimentally observed insulator state reported in Ref.~\onlinecite{yacoby-tbg-fqah-arxiv21}.


\begin{center}
\textbf{\large \V\ B Filling $\nu=7/2$}
\end{center}
To analyze the symmetry of correlated state at $\nu=3$ with doubled moir\'e supercell, in  Fig.~\ref{figs04} we also show the distribution of the expectation value of the valley polarization  (denoted by $\langle \tau_z(\widetilde{\k})\rangle$) within the moir\'e Brillouin zone of the doubled moir\'e supercell at  filling $\nu\!=\!3$ .  Clearly the distribution of the valley polarization in the moir\'e Brillouin zone breaks $C_{3z}$ symmetry, indicating that the density-wave state at $\nu=3$ with doubled moir\'e supercell is a nematic state. 

In the previous calculations, we assume that the ground states at filling 1 and 3 are the SVP density-wave states with doubled moir\'e supercell and with zero Chern numbers. This assumption is justified by the emergence of a diverging instability mode at $M_s$ point at filling 3 as shown in Fig.~\ref{chirpa}(b). Here we also compare the energies between the density-wave states with doubled moir\'e supercell with Chern numbers zero and the states preserving primitive moir\'e translational symmetry with non-zero Chern numbers. At $\nu\!=\!1$ filling, the calculated  energy of the zero-Chern-number density-wave state is lower than  that of the $C\!=\!1$ state (preserving primitive moir\'e translational symmetry) by 0.049\,eV per electron. 
At filling $\nu\!=\!3$, the calculated  energy of the zero-Chern-number density-wave state is lower than  that of the $C\!=\!1$ SVP state (preserving primitive moir\'e translational symmetry) by 0.039\,eV per electron. We see that the energies at both $\nu\!=\!1$ and $\nu\!=\!3$ are significantly lowered by breaking the primitive moir\'e translational symmetry and forming density-wave states with doubled moir\'e supercell. 

\begin{center}
\textbf{\large \V\ C Fillings $\nu=8/3$ and $\nu=11/3$}
\end{center}

\begin{figure}[!htbp]
\includegraphics[width=1.0\textwidth]{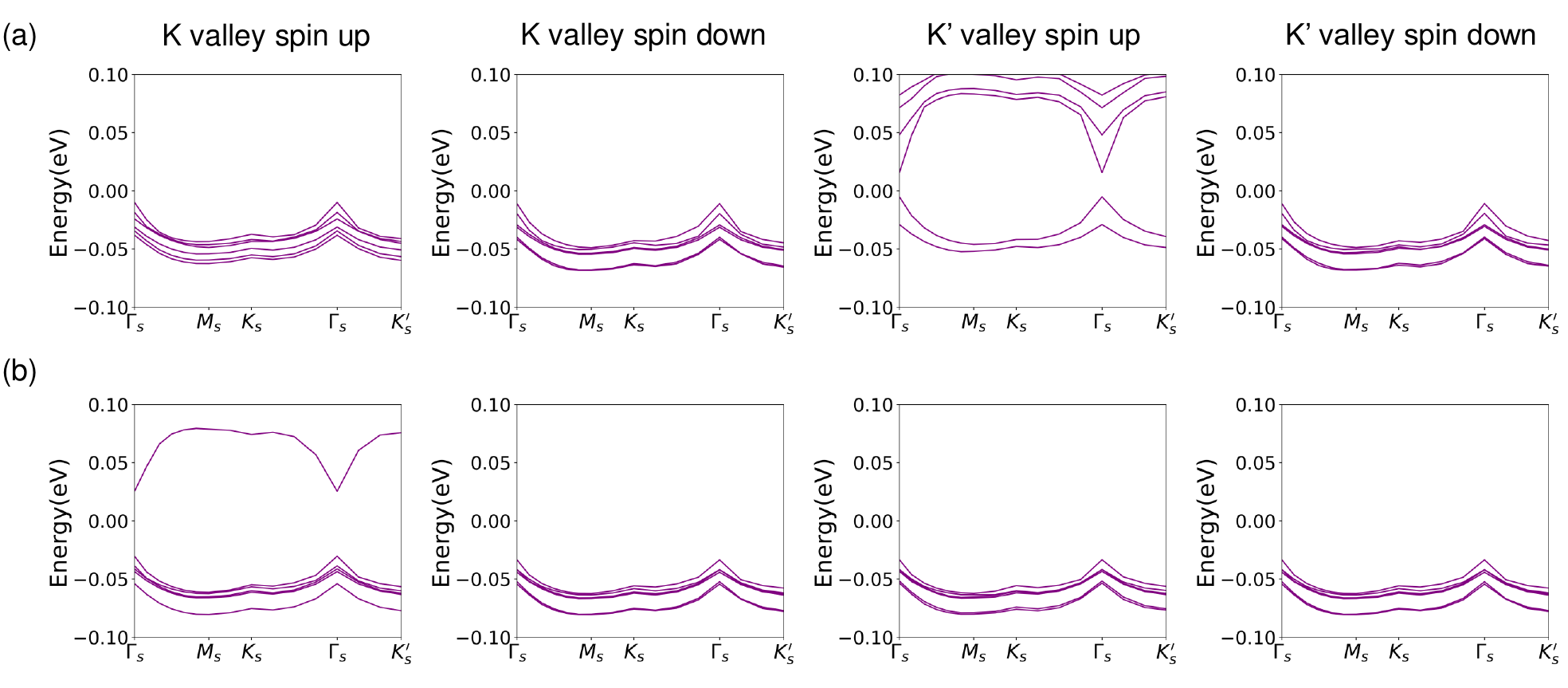}
\caption{~\label{resfig4} The Hartree-Fock energy bands in each valley and spin excluding IVC component at the $\nu\!=\!8/3$ (a) and $\nu\!=\!11/3$ (b) fillings. The Chern number of density wave state at the $\nu\!=\!8/3$ filling is $C\!=\!1$, and the Chern number of DW state at the $\nu\!=\!11/3$ filling is $C\!=\!0$.}
\end{figure}

We continue to discuss the density-wave state at the filling $\nu\!=\!8/3$ with Chern number $C=1$ and that at filling $\nu\!=\!11/3$ with zero Chern number. First, we note that the ground states at $\nu\!=\!8/3$  calculated by the Hartree-Fock+cRPA method are mostly spin-valley-polarized state, with slight mixtures of IVC components; while the ground state at $\nu\!=\!11/3$ is a pure spin-valley polarized state. The ground state at $\nu\!=\!8/3$ is adiabatically connected to a pure spin-valley polarized states without any intervalley coherence. Therefore, in order to better understand the nature of the ground states at $\nu\!=\!8/3$, we temporarily turn off any possible IVC component, and perform restricted Hartree-Fock+cRPA calculations within the subspace of the spin-valley polarized subspace at $\nu\!=\!8/3$. Then each of the energy bands can be resolved with definite valley and spin species. In Fig.~\ref{resfig4}(a) we show the energy bands from by Hartree-Fock+cRPA calculations restricted to the SVP space at 8/3 filling with $\sqrt{3}\times\sqrt{3}$ tripled moir\'e supercell,  where the four panels represent the energy bands from the $K$ valley and spin-up species, $K$ valley and spin-down species, $K'$ valley and spin-up species, and $K'$ valley and  spin-down species, respectively.  We see that only the  $K'$ valley, spin-up bands are partially occupied, with two out of the six bands being occupied; the energy bands of the other three valley-spin species are all fully occupied.  Similarly, in Fig.~\ref{resfig4}(b) we show the Hartree-Fock energy bands at filling $11/3$ with $\sqrt{3}\times\sqrt{3}$ moir\'e supercell, which has a spin-valley-polarized ground state; and we see that only one band from $K$ valley and spin-up species is unoccupied, but all the other bands are occupied. 

It is worthwhile to note that we have also performed unrestricted Hartree-Fock calculations without the cRPA screening effects at $\nu\!=\!8/3$ (with the dielectric constant $\epsilon=7$), and it turns out the ground state is a  fully spin-valley-polarized state, with the occupied electron numbers in each valley and spin species being exactly the same with those presented in Fig.~\ref{resfig4}(a). Thus, we conclude that the slight mixtures of the IVC order at $\nu=8/3$ comes from the screening effects from the remote energy bands.

Now we discuss why the spin-valley polarized states  at filling $\nu\!=\!8/3$ and $\nu\!=\!11/3$ have different Chern numbers. With a realistic choice of the parameters for the continuum model, the ratio $u_0/u_0'\approx0.8$ (see Eq.~(\ref{eq:u}) and discussions therein), which is quite far away from the chiral limit $u_0/u_0'=0$. Breaking chiral symmetry would make the Berry curvature of the flat bands being concentrated near $\Gamma_s$ point \cite{shang-nematic-prr21,ledwith-fci-tbg-prr20,ledwith-strongcoupling-tbg,pablo-tbg-chern-arxiv21}. As a result, one flat band in the primitive moir\'e Brillouin zone would be folded into the Brillouin zone of the $\sqrt{3}\times\sqrt{3}$ supercell, leading to three flat bands. As the Berry curvature is mostly concentrated near $\Gamma_s$ in the primitive moir\'e BZ, one of the three flat bands has  Chern number $\pm 1$, while the other two flat bands have zero Chern numbers. At filling $\nu=8/3$, the two occupied bands from $K'$ valley and spin up species carry Chern numbers 1 and 0, thus the total Chern number of all occupied bands is 1 at 8/3 filling. At filling  $\nu=11/3$, the only unoccupied band from the $K$ valley and spin-up species has zero Chern number, thus Chern number of the occupied bands is also zero. This  explains the origin of the different Chern numbers for  $\nu\!=\!8/3$ and  $\nu\!=\!11/3$, and the calculated Chern numbers are also consistent with recent experimental observations \cite{yacoby-tbg-fqah-arxiv21}.

\begin{figure}[!htbp]
\includegraphics[width=0.4\textwidth]{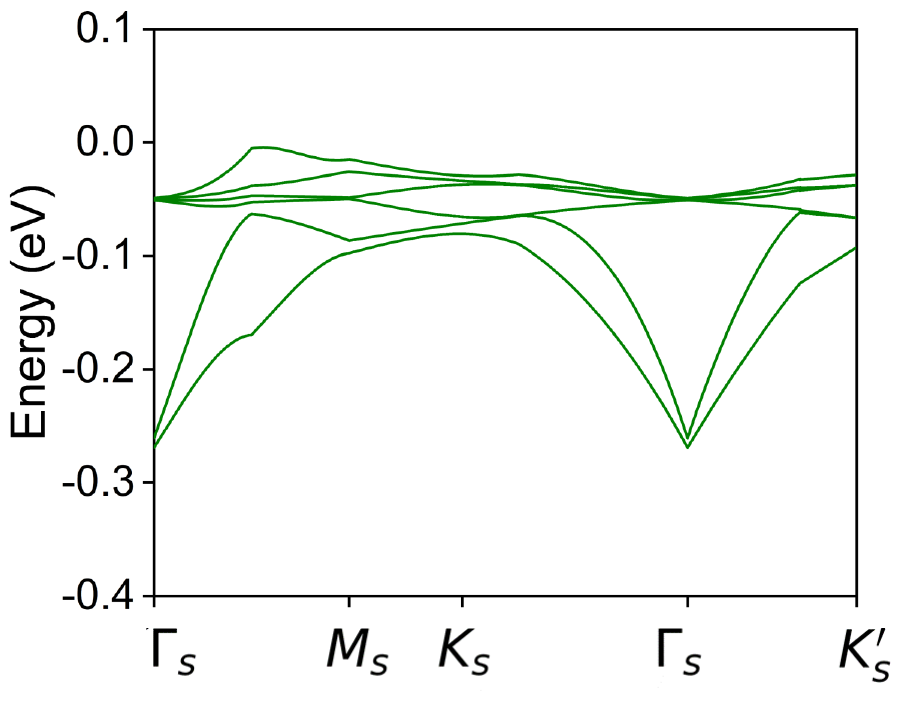}
\caption{~\label{resfig5} The energy bands with only Hartree potential and remote-band potential at the $\nu\!=\!8/3$ filling. Here we use the dielectric constant $\epsilon \!=\!7$ and screening length $d_s\!=\!400$\AA{} in the double-gate screened Coulomb interaction.}
\end{figure}

We are still faced with the problem why the system prefers  valley-spin polarized ground states at fillings $\nu\!=\!8/3$ and  $\nu\!=\!11/3$. 
As discussed in Refs.~\onlinecite{zaletel-tbg-hf-prx20,Bernevig-tbg3-arxiv20,Lian-tbg4-arxiv20}, by virtue of the particle-hole symmetry of the continuum model, the system has an enlarged $U(4)$ symmetry in the flat-band limit (with exactly zero bandwidth), whose generators can be expressed as $\{\tau_x\sigma_y s_a,\tau_y\sigma_y s_a,\tau_z s_a, s_a\}$ ($a=0, x, y, z$) \cite{zaletel-tbg-hf-prx20}, where $\mathbf{\tau}$, $\mathbf{\sigma}$, and $\mathbf{s}$ are Pauli matrices in the valley, sublattice, and spin space, and $s_0$ is the $2\times 2$ identity matrix in the spin space. As a result, in the flat-band limit with broken chiral symmetry (also called "non-chiral, flat limit"), the ground state at the charge neutrality point (filling 2) consists of the degenerate valley polarized (spin-valley polarized) state and the K-IVC state, which are transformed to each other by the $U(4)$ operation. The inclusion of the kinetic energy would further lower the energy of the K-IVC state through the second-order perturbation process \cite{zaletel-tbg-hf-prx20, Lian-tbg4-arxiv20}. 
This is why the K-IVC state is argued to be the ground state at even integer fillings (preserving moir\'e translational symmetry).  On the other hand, at filling $3$ with preserved moir\'e translational symmetry, the ground state is argued to be a fully valley-spin polarized state  \cite{Lian-tbg4-arxiv20}. 
For the case of $\sqrt{3}\times\sqrt{3}$ tripled moir\'e supercell, if the $U(4)$ symmetry of the Hamiltonian is still present in the non-chiral, flat limit, then the above argument should also applies to the even and odd integer fillings of the tripled moir\'e supercell:  filling $\nu=8/3$  corresponds to an even integer filling factor of 8 for the tripled cell, which should have a K-IVC ground state if the $U(4)$ symmetry were still present; while filling $\nu=11/3$ corresponds to an odd integer filling factor of 11 for the tripled moir\'e cell, with one hole per tripled cell, which should have a spin-valley polarized ground state. 

However, the above argument applies only if the $U(4)$ symmetry is still preserved for the $\sqrt{3}\times\sqrt{3}$ tripled moir\'e cell, and the $U(4)$ symmetry arises due to the particle-hole symmetry of the continuum model \cite{Bernevig-tbg3-arxiv20,zaletel-tbg-hf-prx20,hejazi-prb19,jpliu-prb19}. If the primitive moir\'e translational symmetry is broken, the charges would redistribute within the enlarged $\sqrt{3}\times\sqrt{3}$ moir\'e supercell to minimize the Hartree energy, which would break the particle-hole symmetry.
In Fig.~\ref{resfig5}, we present the band structures (plotted in the Brillouin zone of the $\sqrt{3}\times\sqrt{3}$ moir\'e supercell) including  both remote-band Coulomb potentials and the Hartree potentials contributed by the flat-band electrons at filling factor $\nu\!=\!8/3$.  We see that as a result of the  Hartree potential, the energy bands strongly  break particle-hole symmetry, such that the $U(4)$ symmetry is no longer present for the interacting Hamiltonian in the flat-band limit. Therefore, the statement that the ground state at even integer fillings is the K-IVC state no longer applies. In what follows we will show that the ground-state at 8/3 filling is a spin-valley polarized state due to the properties of the interaction form factors.


\begin{figure}[!htbp]
\includegraphics[width=0.6\textwidth]{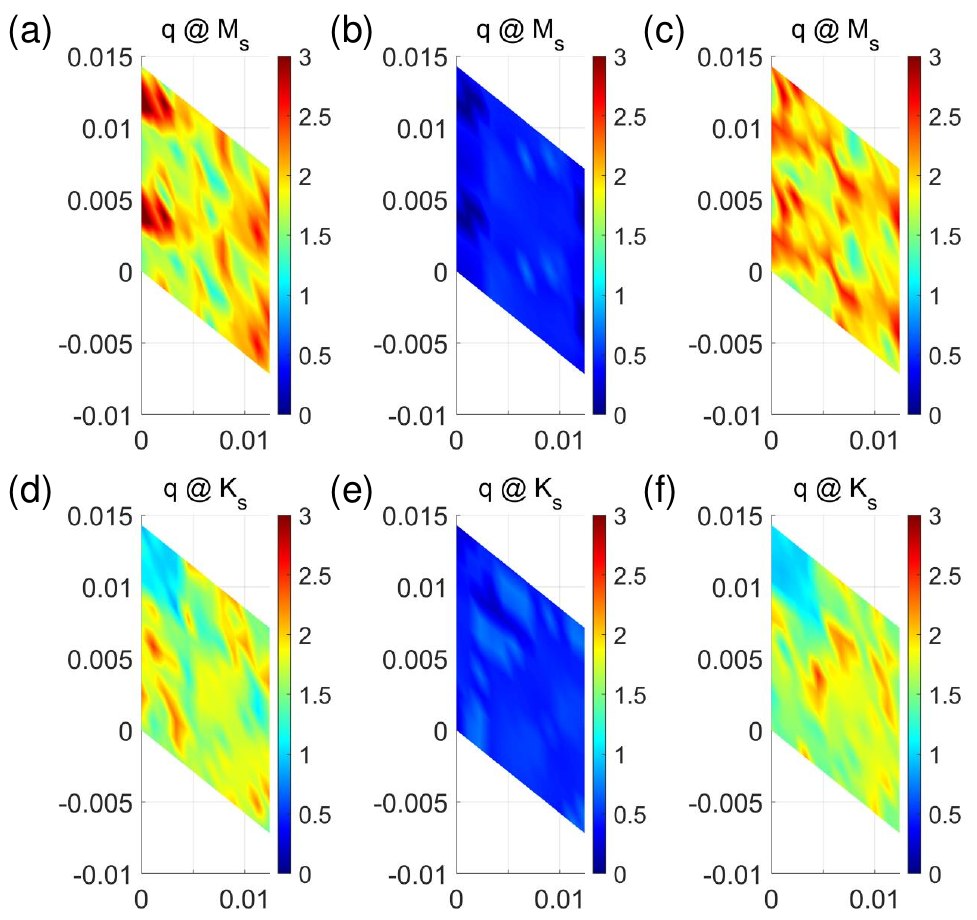}
\caption{~\label{lambda} The $\kt$-dependent distribution of $\rm{tr}_b[\lambda_0(\kt,\qt,\Q =0)\lambda^{\dagger}_0(\kt,\qt,\Q =0)]$ (a,d), $\rm{tr}_b[\lambda_0(\kt,\qt,\Q =0)\lambda^{\dagger}_z(\kt,\qt,\Q =0)]$ (b,e) and $\rm{tr}_b[\lambda_z(\kt,\qt,\Q =0)\lambda^{\dagger}_z(\kt,\qt,\Q =0)]$ (c,f) at the $\nu\!=\!8/3$ filling where Hartree potential and remote-band Coulomb potential are taken into consideration. In the (a-c), $\qt=\G_1/2$. In the (d-f), $\qt=\G_1/3+2\G_2/3$. We use the dielectric constant $\epsilon \!=\!7$ and screening length $d_s\!=\!400$\AA{} in the Coulomb interaction. In the calculations, we take $12\times 12$ k-point mesh in the triple moir\'e supercell.}
\end{figure}

First, we define the form factor
\begin{equation}
\lambda _{\mu \sigma m,\mu \sigma n}(\kt,\qt,\Q)=\sum _{\alpha \G}C^*_{\mu \alpha \G+\Q,n\kt + \qt}C_{\mu \alpha \G, m\kt}
\label{eq:lambda}
\end{equation}
where $C_{\mu \alpha \G, m\kt}$ is the non-interacting wavefunction, and $\mu$, $\sigma$, and $\alpha$ refer to the valley, spin, and layer/sublattice degrees of freedom.

First we discuss  the constraint on the form factor from the particle-hole symmetry $\mathcal{P}=i\tau_xl_y\sigma_x$ (in some literatures this is known as the $\mathcal{P}\mathcal{T}$ symmetry) \cite{song-tbg-prl19,hejazi-prb19,jpliu-prb19,zaletel-tbg-hf-prx20,shang-nematic-prr21,Bernevig-tbg3-arxiv20}, where $\mathbf{\tau}$, $\mathbf{l}$ and $\mathbf{\sigma}$ denote Pauli matrices in the valley, layer and sublattice space. 
The particle-hole symmetry is manifested as $\mathcal{P} H^{\mu}(\k)\mathcal{P}^{-1}=-H^{-\mu}(\k)$, where $H^{+/-\mu}(\k)$ refers to the continuum Hamiltonian of valley $\pm\mu$. Then we can obtain the following relationship between the eigenstates from the $+/-\mu$ valley 
\begin{align}
&\mathcal{P}\vert\psi_{\mu,n\kt}\rangle=\vert \psi_{-\mu,n\kt} \rangle e^{i\theta_{\mu}(\kt)}\;\nn
&E_{\mu,n\kt}=-E_{-\mu,n\kt}\;,
\label{eq:pgauge}
\end{align}
 where $\vert\psi_{\mu,n\kt} \rangle=\sum_{\alpha\G} C_{\mu\alpha\G,n\kt} |\kt+\G,\mu\alpha \rangle$ is the non-interacting Bloch eigenstate of the $n$th band from valley $\mu$ at moir\'e wavevector $\kt$, $E_{\mu,n\kt}$ is the corresponding eigenenergy, $\alpha$ refers to the layer and sublattice indices, and $\mathbf{G}$ refers to the moir\'e reciprocal vector.
Here $\theta_{\mu}(\kt)$ is a gauge freedom and can be fixed to zero. Eq.~(\ref{eq:pgauge}) can be explicitly written as 
\begin{align}
\mathcal{P}\vert\psi_{\mu,n\kt}\rangle &= \sum_{\alpha\G}C_{\mu\alpha\G,n\kt}\,i\sum_{\alpha 'mu'}(\tau _x)_{\mu '\mu}(l_y\sigma_x)_{\alpha '\alpha}|\kt+\G,\mu'\alpha' \rangle\nn
&=\sum_{\alpha '\G}\left[\sum_{\alpha}\,C_{\mu\alpha\G,n\kt}\,(il_y\sigma _x)_{\alpha '\alpha}\right] |\kt+\G,-\mu\,\alpha' \rangle\nn
&=\vert\psi_{-\mu,n\kt}\rangle=\sum_{\alpha '\G}C_{-\mu\alpha '\G,n\kt} |\kt+\G,-\mu\,\alpha' \rangle\;.
\end{align}
It follows that the non-interacting wavefunction can be fixed by the following particle-hole gauge
\begin{equation}
\sum_{\alpha '} C_{\mu\alpha '\G,n\kt}(il_y\sigma_x)_{\alpha\alpha '}=C_{-\mu\alpha\G,n\kt}
\end{equation}
Then the form factors of the opposite valleys have to obey the following relationship,
\begin{align}
\lambda _{\mu\sigma m,\mu\sigma n}(\kt,\qt,\Q)&=\sum_{\alpha\G}C^*_{\mu\alpha\G+\Q,n\kt+\qt}C_{\mu\alpha\G,m\kt}\nn
&=\sum_{\alpha\G}\sum_{\alpha _1}(il_y\sigma _x)^*_{\alpha\alpha _1}C^*_{-\mu\alpha _1\G+\Q,n\kt+\qt}\sum_{\alpha _2}(il_y\sigma _x)_{\alpha \alpha _2}C_{-\mu\alpha _2\G,m\kt}\nn
&=\sum _{\alpha _1\alpha _2\G}C^*_{-\mu\alpha _1\G+\Q,n\kt+\qt}C_{-\mu\alpha _2\G,m\kt}\delta_{\alpha _1\alpha _2}\nn
&=\lambda _{-\mu\,\sigma m,-\mu\,\sigma n}(\kt,\qt,\Q)\;.
\label{eq:form-P}
\end{align}

Then we define the density operator $\Delta_{\mu\sigma n,\mu '\sigma 'm'}(\kt)$ as 
\begin{equation}
\Delta_{\mu\sigma n,\mu '\sigma ' m'}(\kt)=\langle \hat{c}^{\dagger}_{\mu\sigma n,\kt}\hat{c}_{\mu '\sigma 'm',\kt} \rangle\;,
\end{equation}
which can be rewritten as
\begin{equation}
\hat{\Delta}_{\mu\sigma n,\mu '\sigma ' m'}(\kt)=\sum_{\substack{a=0,x,y,z\\ b=0,z}}\tau_a s_b \hat{Q}^{ab}\;,
\label{eq:Delta}
\end{equation}
where $\hat{Q}$ is a $6\times6$ matrix defined with the flat-band basis in the triple moir\'e supercell (there are six flat bands per spin per valley for tripled moir\'e supercell). The density operator satisfies $\rm{tr}[\hat{\Delta}]=\nu$ in which $\nu$ is filling factor. Moreover, as we are considering insulator state, the density operator is also a projector onto the occupied subspace, which satisfies $\hat{\Delta}^2=\hat{\Delta}$.

With the above considerations, the Fock energy can be expressed in terms of the form factors and the density matrices as
\begin{align}
E^F &= -\frac{1}{2N_s}\sum_{\kt \qt}\sum _{\Q} V(\qt+\Q)\sum _{\substack{\mu \mu ' \sigma \sigma ' \\mnm'n'}}\Delta _{\mu ' \sigma ' n', \mu \sigma m}(\kt)\lambda _{\mu \sigma m, \mu \sigma n}(\kt, \qt, \Q)\Delta _{\mu  \sigma  n, \mu '\sigma 'm'}(\kt+\qt)\lambda ^{\dagger}_{\mu '\sigma 'm', \mu '\sigma 'n'}(\kt, \qt, \Q)\nn
&=-\frac{1}{2N_s}\sum_{\kt \qt}\sum _{\Q} V(\qt+\Q)\rm{tr}\left[ \hat{\Delta}(\kt)\hat{\lambda}(\kt,\qt,\Q)\hat{\Delta}(\kt+\qt)\hat{\lambda}^{\dagger}(\kt,\qt,\Q) \right]
\label{eq:fock-trace}
\end{align}
It should be noted that the form factor $\hat{\lambda}(\kt,\qt,\Q)$, which is a $24\times 24$ matrix as defined in Eq.~(\ref{eq:lambda}),  can be written in block diagonal form in the valley-spin space,
\begin{equation}
\hat{\lambda}(\kt,\qt,\Q)=\begin{pmatrix}
\hat{\lambda}_{+,+}(\kt,\qt,\Q) & 0 &0 &0 \\
0 & \hat{\lambda}_{+,+}(\kt,\qt,\Q) &0 &0 \\
0 & 0 &\hat{\lambda}_{-,-}(\kt,\qt,\Q) &0 \\
0 &0 &0 & \hat{\lambda}_{-,-}(\kt,\qt,\Q) \\
\end{pmatrix}\;.
\end{equation}
Here $+/-$ refers to the $K/K'$ valley, and the form factor is spin independent For clarity, the form factor can be re-written as
\begin{equation}
\hat{\lambda}(\kt,\qt,\Q)=\hat{\lambda}_0(\kt,\qt,\Q)\mathbb{I}_{4\times 4}+\hat{\lambda}_z(\kt,\qt,\Q)\tau_z\otimes s_0
\label{eq:lambda-matrix}
\end{equation}
where
\begin{align}
\hat{\lambda}_0(\kt,\qt,\Q)=[\hat{\lambda}_{+,+}(\kt,\qt,\Q)+\hat{\lambda}_{-,-}(\kt,\qt,\Q)]/2, \\
\hat{\lambda}_z(\kt,\qt,\Q)=[\hat{\lambda}_{+,+}(\kt,\qt,\Q)-\hat{\lambda}_{-,-}(\kt,\qt,\Q)]/2,
\end{align}
are both $6\times 6$ matrices in the flat-band basis, $\mathbb{I}_{4\times 4}$ is the $4\times 4$ identity matrix in the valley-spin space, $\tau_z$ is the third Pauli matrix in valley space, and $s_0$ is the identity matrix in spin space.
Plugging Eq.~(\ref{eq:lambda-matrix}) and Eq.~(\ref{eq:Delta}) into Eq.~(\ref{eq:fock-trace}), the Fock energy can be further expressed as
\begin{align}
\label{eq:EFock}
E^F =& -\frac{1}{N_s} \sum_{\kt,\qt}\sum_{\Q}V(\qt+\Q)\sum_{\substack{a,b,a',b'}}\Big(\,\rm{tr}_{vs}[\tau_a\,s_b\,\mathbb{I}_{4\times4}\,\tau_{a'}\,s_{b'}\,\mathbb{I}_{4\times 4})\rm{tr}_{b}[\hat{Q}^{ab}\,\hat{\lambda_0}\,\hat{Q}^{a'b'}\hat{\lambda}_0^{\dagger}]\nn
&+ \rm{tr}_{vs}[\tau_a\,s_b\,\tau_z\,\tau_{a'}\,s_{b'}\,\mathbb{I}_{4\times 4}]\,\rm{tr}_{b}(\hat{Q}^{ab}\,\hat{\lambda}_z\,\hat{Q}^{a'b'}\,\hat{\lambda}^{\dagger}_0]\nn
&+ \rm{tr}_{vs}[\tau_a\,s_b\,\mathbb{I}_{4\times4}\,\tau_{a'}\,s_{b'}\,\tau_z]\rm{tr}_{b}[\hat{Q}^{ab}\,\hat{\lambda_0}\,\hat{Q}^{a'b'}\,\hat{\lambda}^{\dagger} _z]\nn
&+ \rm{tr}_{vs}[\tau_a\,s_b\,\tau_z\,\tau_{a'}\,s_{b'}\,\tau_z]\rm{tr}_{b}[\hat{Q}^{ab}\,\lambda_z\,\hat{Q}^{a'b'}\,\lambda^{\dagger}_z]\,\Big)\;,
\end{align}
where $\rm{tr}_{vs}[...]$ and $\rm{tr}_{b}[...]$ denote taking the partial trace of the matrix in the valley-spin subspace and the flat-band subspace respectively. First, we note that for any $\kt$-independent order parameters, the first term can  be re-written as $\rm{tr}[\hat{\Delta}(\kt)\,\hat{\Delta}(\kt+\qt)]=\rm{tr}[\hat{\Delta}(\kt)]=\nu$, which is only dependent on the filling factor. Thus the first term contributes to the same Fock energy for any type of $\kt$-independent order parameters. The second and third terms would favour a strongly $\kt$-dependent IVC order parameter or $\kt$-independent valley polarized order parameter, which make $\rm{tr}_{vs}[\tau^as^b\mathbb{I}\tau^{a'}s^{b'}\tau_z]\!>\!0$ or $\rm{tr}_{vs}[\tau^as^b\tau_z\tau^{a'}s^{b'}\mathbb{I}]\!>\!0$. As for the last term, clearly it favours a spin and/or valley polarized state due to Cauchy-Schwarz inequality \cite{shang-nematic-prr21,zhang-tbmg-prl22}  
\begin{equation}
\rm{tr}_{vs}[\tau_a\,s_b\,\tau_z\,\tau_{a'}\,s_{b'}\,\tau_z]\leq \rm{tr}_{vs}[\tau_a\,s_b]\rm{tr}_{vs}[\tau_z\,\tau_{a'}\,s_{b'}\tau_z]\;.
\label{eq:cauchy}
\end{equation}
The equality condition is satisfied if and only if
\begin{equation}
\tau_a\,s_b=\tau_z\,\tau_{a'}\,s_{b'}\tau_z\;.
\end{equation}
This gives us three spin-valley polarized order parameters $\tau_z$, $s_z$, and $\tau_z s_z$ which satisfy the above equation. However, if there is $\mathcal{P}$ symmetry,  the form factor $\hat{\lambda}$ should satisfy Eq.~(\ref{eq:form-P}), thus $\hat{\lambda}_z=0$, and the last term in Eq.~(\ref{eq:EFock}), thus the system not necessarily favors a spin-valley polarized state.  For the case of magic-angle TBG, further analysis reveals that the K-IVC state is the ground state at filling 0 \cite{zaletel-tbg-hf-prx20, shang-nematic-prr21}.

However, as shown in Fig.~\ref{resfig5}, particle-hole symmetry is broken for the tripled moir\'e cell at 8/3 filling due to the Hartree-potential, thus Eq.~(\ref{eq:form-P}) no longer holds and $\hat{\lambda}_z\neq 0$. Now we  carefully check the amplitudes of each of the four terms in Eq.~(\ref{eq:EFock}), trying to find out the dominant terms and to determine the ground state at 8/3 filling. In particular, we have calculated the $\kt$ dependence of the form factors $\rm{tr}_b[\lambda_0(\kt,\qt,\Q=0)\lambda^{\dagger}_0(\kt,\qt,\Q=0)]$, $\rm{tr}_b[\lambda_0(\kt,\qt,\Q=0)\lambda^{\dagger}_z(\kt,\qt,\Q=0)]$, and $\rm{tr}_b[\lambda_z(\kt,\qt,\Q=0)\lambda^{\dagger}_z(\kt,\qt,\Q=0)]$ for the tripled cell at 8/3 filling (including Hartree potentials) at $\qt=\mathbf{G}_1/2$ and $\qt=\mathbf{G}_1/3+2\mathbf{G}_2/3$.
 As shown in  Fig.~\ref{lambda}, the amplitude of $\rm{tr}_b[\lambda_0(\kt,\qt,\Q=0)\lambda^{\dagger}_z(\kt,\qt,\Q=0)]$ is much smaller than those of $\rm{tr}_b[\lambda_0(\kt,\qt,\Q)\lambda^{\dagger}_0(\kt,\qt,\Q)]$ and $\rm{tr}_b[\lambda_z(\kt,\qt,\Q)\lambda^{\dagger}_z(\kt,\qt,\Q)]$. Thus the dominant term in the Fock energy is the first term and last term in Eq.~(\ref{eq:EFock}), which favours spin-valley polarized states. Similar argument can also be applied to the situation at filling 11/3. In summary, the broken particle-hole symmetry at 8/3 and 11/3 fillings due to the Hartree potentials in the tripled moir\'e supercell is the essential reason for the spin-valley polarized ground states at these fillings. 

\vspace{12pt}
\begin{center}
\textbf{\large \VI\ Generalized susceptibility calculations}
\end{center}

\begin{figure}[!htbp]
\includegraphics[width=1.0\textwidth]{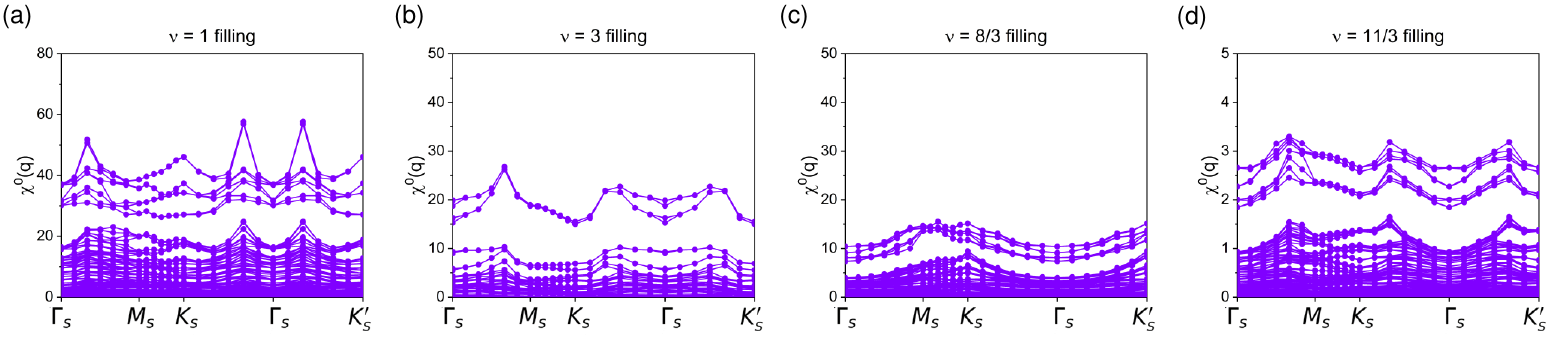}
\caption{~\label{chi0} The bare susceptibility at (a) $\nu\!=\!1$,(b) $\nu\!=\!3$, (c) $\nu\!=\!8/3$ and (d) $\nu\!=\!11/3$ fillings. Here we adopt the 48$\times$48 k-points mesh and remote band Hartree-Fock potential. We use the dielectric constant $\epsilon \!=\!10$ and screening length $d_s\!=\!300$\AA{} in the Coulomb interaction at $\nu\!=\!1$, 3 and 8/3 fillings. At $\nu\!=\!11/3$ filling, we use the dielectric constant $\epsilon \!=\!7$ to search the obvious diverging modes.}
\end{figure}

\begin{figure}[!htbp]
\includegraphics[width=1.0\textwidth]{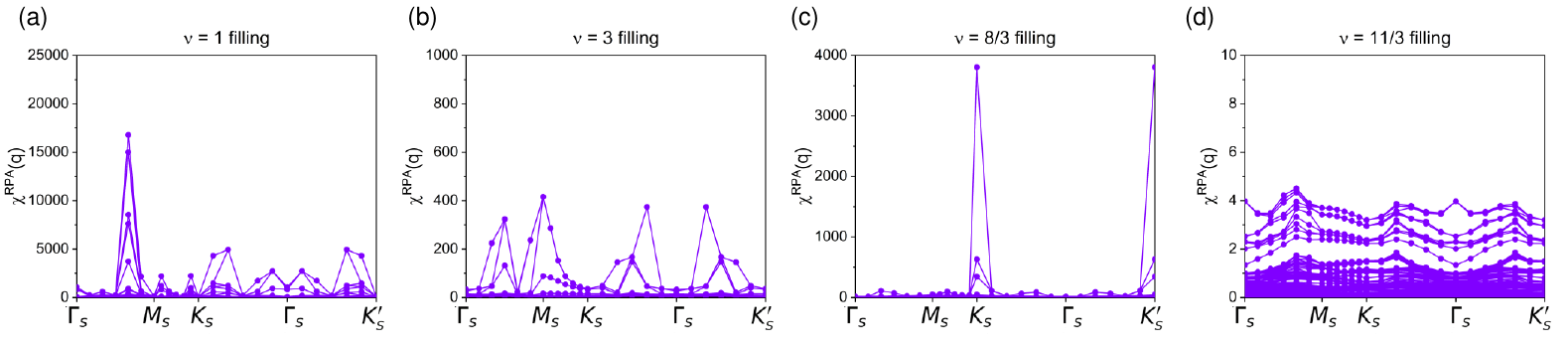}
\caption{~\label{chirpa} The RPA susceptibility at (a) $\nu\!=\!1$,(b) $\nu\!=\!3$, (c) $\nu\!=\!8/3$ and (d) $\nu\!=\!11/3$ fillings. Here we adopt the 48$\times$48 k-points mesh. And the remote band potential is also included in the RPA calculations. We use the dielectric constant $\epsilon \!=\!10$ and screening length $d_s\!=\!300$\AA{} in the Coulomb interaction at $\nu\!=\!1$, 3 and 8/3 fillings. At $\nu\!=\!11/3$ filling, we use the dielectric constant $\epsilon \!=\!7$ to search the obvious diverging modes.}
\end{figure}

In the previous Hartree-Fock calculations with broken moir\'e translational symmetry, we have made some specific choices of the moir\'e supercells at different integer and fractional filling factors. Namely, at filling 1, 3, and 7/2, we have chosen a doubled moir\'e supercell;  at filling 8/3, we have considered a $\sqrt{3}\times\sqrt{3}$ tripled moir\'e supercell; while at filling 11/3, we have considered both $\sqrt{3}\times\sqrt{3}$ and $3\times 1$ moir\'e supercells. Such choices of density-wave states can be justified by directly calculating the moir\'e wavevector dependence of the instability modes for the Fermi surfaces at different filling factors. In this section, we try to find  the leading Fermi-surface instability modes by calculating the generalized susceptibility tensor at some moir\'e wavevector $\qt$ under random phase approximation (RPA). To be specific, we define the bare susceptibility tensor in the original valley-spin-layer-sublattice basis  of the continuum model as:
\begin{align}
&\chi^{0}_{\mu\sigma\alpha\mathbf{G},\mu'\sigma'\beta\mathbf{G}+\mathbf{Q}\,;\,\mu\sigma\alpha'\mathbf{G}',\mu'\sigma'\beta'\mathbf{G}'+\mathbf{Q}}(\widetilde{\mathbf{q}},i\nu_n)\nn
=&-\frac{1}{\beta}\sum_{\i\omega_n}\int \frac{dk^2}{(2\pi)^2}\,G^{0}_{\mu\sigma\alpha\mathbf{G},\mu\sigma\alpha'\mathbf{G}'}(i\omega_n,\widetilde{\mathbf{k}})\,G^{0}_{\mu'\sigma'\beta'\,\mathbf{G}'+\mathbf{Q},\mu'\sigma'\beta\;\mathbf{G}+\mathbf{Q}}(i\omega_n+i\nu_n,\widetilde{\mathbf{k}}+\widetilde{\mathbf{q}})\;
\label{eq:chi0}
\end{align}
where $\mu,\mu'$ are the valley indices, $\sigma,\sigma'$ are the spin indices, $\alpha,\beta,\alpha',\beta'$ are the layer and sublattice indices, $\mathbf{G}, \mathbf{G}', \mathbf{Q}$ are the moir\'e reciprocal lattice vectors, $\widetilde{\mathbf{k}},\widetilde{\mathbf{q}}$ are the wavevectors within the moir\'e Brillouin zone. $\i\omega_n, i\nu_n$ are the Fermionic and Bosonic Matsubara frequencies, and $\beta=1/k_B T$, with $k_B$ denoting the Boltzmann constant and $T$ denoting the temperature. $G^{0}_{\mu\sigma\alpha\mathbf{G},\mu\sigma\alpha'\mathbf{G}'}(i\omega_n,\widetilde{\mathbf{k}})$ denotes the non-interacting single-particle Green's function expressed in the original basis of the continuum model:
\begin{equation}
G^{0}_{\mu\sigma\alpha\mathbf{G},\mu\sigma\alpha'\mathbf{G}'}(i\omega_n,\widetilde{\mathbf{k}})=\sum_{n\in \rm{flat}} \frac{C^{*}_{\mu\alpha\mathbf{G},n\kt}\,C_{\mu\alpha'\mathbf{G}',n\kt}}{i\omega_n-E_{\mu,n\widetilde{\mathbf{k}}}}
\label{eq:green}
\end{equation}
where the $C_{\mu\alpha\mathbf{G},n\kt}$ denotes the non-interacting wavefunction at $\widetilde{\mathbf{k}}$, and $n$ is the band index. Since the Coulomb interaction effects are most prominent for the flat bands, in Eq.~(\ref{eq:green}), and we are interested in the instability modes driven by quantum fluctuations of the flat bands, the summation of the band index $n$ in Eq.~(\ref{eq:green}) is restricted   to the flat-band subspace. $E_{\mu,n\kt}$ denotes the flat-band dispersion \textit{including the Coulomb potentials from the occupied remote energy bands}. It is worthwhile to note that, the Coulomb potentials from the remote energy bands make the flat bands much more dispersive, with a bandwidth $\sim 50\,$meV as shown in  Fig.~1(d) of main text, which somehow justifies the RPA treatment in calculating the generalized susceptibility tensor.
Plugging Eq.~(\ref{eq:green}) into Eq.~(\ref{eq:chi0}), and carrying out the summation over Matsubara frequency, one obtains
\begin{align}
&\chi^{0}_{\mu\sigma\alpha\mathbf{G},\mu'\sigma'\beta\;\mathbf{G}+\mathbf{Q}\;;\;\mu\sigma\alpha'\mathbf{G}',\mu'\sigma'\beta'\;\mathbf{G}'+\mathbf{Q}}(\widetilde{\mathbf{q}},i\nu_n)\;\nn
=&\int\frac{dk^2}{(2\pi)^2}\sum_{n,m}\,C^{*}_{\mu\alpha\mathbf{G},n\kt}\,C_{\mu\alpha'\mathbf{G}',n\kt}\,C^{*}_{\mu'\beta'\;\mathbf{G}'+\mathbf{Q},m\kt+\qt}\,C_{\mu'\beta\;\mathbf{G}+\mathbf{Q},m\kt+\qt}\,\frac{f(E_{\mu,n\widetilde{\k}})-f(E_{\mu',m\widetilde{\k}+\widetilde{\mathbf{q}}})}{E_{\mu',m\widetilde{\k}+\widetilde{\mathbf{q}}}-E_{\mu,n\widetilde{\k}}-i\nu_n}
\label{eq:chi0}
\end{align}
Such a susceptibility tensor defined in the original basis characterizes the intrinsic Fermi-surface fluctuations in the valley-spin-layer-sublattice space. In order to describe the possible spontaneous symmetry-breaking states, here we only consider the zero-frequency susceptibility with $i\nu_n\!=\!0$, and keep the full moir\'e wavevector dependence: 
\begin{equation}
\chi^{0}_{\mu\sigma\alpha\mathbf{G},\mu'\sigma'\beta\;\mathbf{G}\;;\;\mu\sigma\alpha'\mathbf{G}',\mu'\sigma'\beta'\;\mathbf{G}'}(\widetilde{\mathbf{q}},\mathbf{Q})\equiv\chi^{0}_{\mu\sigma\alpha\mathbf{G},\mu'\sigma'\beta\;\mathbf{G}+\mathbf{Q}\;;\;\mu\sigma\alpha'\mathbf{G}',\mu'\sigma'\beta'\;\mathbf{G}'+\mathbf{Q}}(\widetilde{\mathbf{q}},i\nu_n\!=\!0)\;.
\end{equation}
Electron-electron Coulomb interactions may greatly enhance the susceptibility tensor and drive a second-order phase transition via spontaneous symmetry breaking in the valley-spin-layer-sublattice space. In particular, we consider the dominant intravalley Coulomb interactions as given by Eq.~(\ref{eq:h-intra}). The two-particle Coulomb scattering processes involve  those from both direct  and  exchange Coulomb interactions, which can be written in  matrix form as:
\begin{align}
&\mathbb{U}(\widetilde{\mathbf{q}},\mathbf{Q})_{\mu_{\alpha}\sigma_{\alpha}\alpha\mathbf{G},\mu_{\beta}\sigma_{\beta}\beta\mathbf{G}\;;\;\mu_{\alpha}'\sigma_{\alpha}'\alpha'\mathbf{G}',\mu_{\beta}'\sigma_{\beta}'\beta'\mathbf{G}'}\;\nn
=&V(\vert\widetilde{\mathbf{q}}+\mathbf{Q}\vert)\,\delta_{\mu_{\alpha}\mu_{\beta}}\delta_{\sigma_{\alpha}\sigma_{\beta}}\delta_{\alpha\beta}-V(\vert\widetilde{\mathbf{k}}+\mathbf{G}-\widetilde{\mathbf{k}}'-\mathbf{G}'\vert)\,\delta_{\mu_{\alpha}\mu_{\alpha}'}\delta_{\mu_{\beta}\mu_{\beta}'}\delta_{\sigma_{\alpha}\sigma_{\alpha}'}\delta_{\sigma_{\beta}\sigma_{\beta}'}\delta_{\alpha\alpha'}\delta_{\beta\beta'}\delta_{\mu_{\alpha}'\mu_{\beta}'}\delta_{\sigma_{\alpha}'\sigma_{\beta}'}\delta_{\alpha'\beta'}\;\nn
\approx &V(\vert\widetilde{\mathbf{q}}+\mathbf{Q}\vert)\,\delta_{\mu_{\alpha}\mu_{\beta}}\delta_{\sigma_{\alpha}\sigma_{\beta}}\delta_{\alpha\beta}-V(\vert\mathbf{G}-\mathbf{G}'\vert)\,\delta_{\mu_{\alpha}\mu_{\alpha}'}\delta_{\mu_{\beta}\mu_{\beta}'}\delta_{\sigma_{\alpha}\sigma_{\alpha}'}\delta_{\sigma_{\beta}\sigma_{\beta}'}\delta_{\alpha\alpha'}\delta_{\beta\beta'}\delta_{\mu_{\alpha}'\mu_{\beta}'}\delta_{\sigma_{\alpha}'\sigma_{\beta}'}\delta_{\alpha'\beta'}
\label{eq:coulomb-matrix}
\end{align}
Again, $\{\mu_{\alpha},\mu_{\beta},\mu_{\alpha'},\mu_{\beta '}\}$, $\{\sigma_{\alpha},\sigma_{\beta},\sigma_{\alpha'},\sigma_{\beta '}\}$, and $\{\alpha,\beta,\alpha',\beta'\}$ denote the valley, spin, and layer/sublattice indices respectively, and $V(\mathbf{q})$ is the double-gate screened Coulomb interaction as shown in Eq.~(\ref{eq:double-gate}). It is important to note that, in the last line of Eq.~(\ref{eq:coulomb-matrix}), we have made an approximation that the amplitude of the exchange Coulomb interaction is only dependent on the transfer of reciprocal vector  $\mathbf{G}-\mathbf{G}'$, neglecting the transfer of the moir\'e wavevector $\widetilde{\mathbf{k}}-\widetilde{\mathbf{k}}'$ within the moir\'e Brillouin zone. This is an excellent approximation given the small size of the moir\'e Brillouin zone around the magic angle. Then we calculate the interaction-renormalized generalized susceptibility tensor with random phase approximation. By virtue of the approximation made in Eq.~(\ref{eq:coulomb-matrix}), the RPA susceptibility can be written in a succinct matrix form:
\begin{equation}
\hat{\chi}^{\rm{RPA}}(\widetilde{\mathbf{q}},\mathbf{Q})=\hat{\chi}^{0}(\qt,\mathbf{Q})\,\cdot\,(1+\mathbb{U}(\qt,\mathbf{Q})\cdot\hat{\chi}^{0}(\widetilde{q},\mathbf{Q}))^{-1}
\end{equation}
where $\hat{\chi}^{0}(\widetilde{q},\mathbf{Q}))$ is the matrix of bare susceptibility, whose matrix element is given in Eq.~(\ref{eq:chi0}), and $\hat{\chi}^{\rm{RPA}}(\widetilde{\mathbf{q}},\mathbf{Q})$ is the RPA susceptibility tensor defined in the same basis as  $\hat{\chi}^{0}(\qt,\mathbf{Q}))$ and $\mathbb{U}(\widetilde{\mathbf{q}},\mathbf{Q})$. In the end, we sum over all the transferred reciprocal vectors $\mathbf{Q}$, and define the RPA susceptibility at a moir\'e wavevector $\widetilde{\mathbf{q}}$ (within moir\'e Brillouin zone) as:
\begin{equation}
\hat{\chi}^{\rm{RPA}}(\widetilde{\mathbf{q}})=\sum_{\mathbf{Q}}\,\hat{\chi}^{\rm{RPA}}(\widetilde{\mathbf{q}},\mathbf{Q})
\label{eq:chiRPA}
\end{equation}
which captures the Fermi-surface quantum fluctuations in the valley-spin-layer-sublattice space contributed by the flat bands.  One can diagonalize $\hat{\chi}^{\rm{RPA}}(\widetilde{\mathbf{q}})$ at each moir\'e wavevector $\widetilde{\mathbf{q}}$, and any instability modes with diverging eigenvalues would indicate the tendency of forming a symmetry-breaking density-wave state with wavevector $\widetilde{\mathbf{q}}$. The  eigenvectors of the diverging modes would correspond to the order parameters of the possible density-wave states.

We present the eigenvalues of the bare susceptibilities and the interaction-renormalized RPA susceptibilities in the Figs.~\ref{chi0} and Fig.~\ref{chirpa}. It can be noted that electron-electron Coulomb interactions play an important role in driving the system to the density-wave phases at the $\nu\!=\!1$, 3, and 8/3 fillings. Especially, the leading instability mode with diverging eigenvalue is located at the $M_s$ point at filling 3, where the corresponding eigenmode is a spin-valley polarized mode that is consistent with the Hartree-Fock calculations. This justifies the choice of a doubled moir\'e supercell at this filling.  Turning to the 8/3 filling, the leading instability mode is located at the $K_s$ point, which implies that the system favours a CDW state with $\sqrt{3}\times\sqrt{3}$ tripled moir\'e supercell. The leading eigenmodes at 8/3 filling at $K_s$ point involve both IVC modes and spin-valley polarized modes, and the latter turn out to be the ground state according to the Hartree-Fock calculations.  Furthermore, we have also calculated the density-wave state with $3\times 1$ tripled moir\'e supercell at 8/3 filling, and found that the ground state energy of such a period-3 stripe state is higher than that  of the $\sqrt{3}\times\sqrt{3}$ tripled moir\'e supercell by 5.2\,meV per electron at 8/3 filling.

Turning to filling 1, the leading instability mode is located somewhere between the $\Gamma _s$ and $M_s$ points with the eigenvector involving both the spin-valley polarized mode and some of the ``nematic" modes ($\tau_z\sigma_x$,$\sigma_y$), which may lead to the ``incommensurate Kekul\'e spiral state" (IKS) with incommensurate wavevector \cite{zaletel-tbg-kekule-PRX21}. Since the IKS state has been extensively discussed previously, here we explore another simple choice of density-wave state at filling 1 that could lead to an insulator state, which is simply to double the moir\'e supercell. 
We have compared  the energies of the $C$=0 (with doubled cell) and $C$=1 (with primitive cell) state at $\nu\!=\!1$ filling, and surprisingly found that the calculated energy of the zero-Chern-number density-wave state is lower than that of the $C\!=\!1$ state (preserving primitive moir\'e translational symmetry) by 15.15\,meV per electron.  This indicates that the density wave state with doubled moir\'e supercell is possible candidate for the ground state at filling 1.
 
 At filling $\nu\!=\!3$, the calculated  energy of the $C\!=\!0$ density-wave state is lower than  that of the $C\!=\!1$ SVP state (preserving primitive moir\'e translational symmetry) by 14.07\,meV per electron. We see that the energies at both $\nu\!=\!1$ and $\nu\!=\!3$ are significantly lowered by breaking the primitive moir\'e translational symmetry and forming density-wave states with doubled moir\'e supercell.

As for filling 11/3, we do not find prominent instability modes along the high-symmetry path. In order to obtain an insulator state at $\nu=11/3$, one has to triple the moir\'e supercell. We have considered two types of tripled supercells: the $\sqrt{3}\times\sqrt{3}$ supercell and the $3\times 1$ stripe supercell. Our Hartree-Fock calculations indicate the ground-state energy of the $\sqrt{3}\times\sqrt{3}$ supercell is lower than that of the $3\times 1$ supercell by 3.9\,meV per electron. Therefore, in our work a $\sqrt{3}\times\sqrt{3}$ supercell is adopted at filling 11/3, and the ground state is predicted to be a spin-valley polarized state with zero Chern number.

\vspace{12pt}
\begin{center}
\textbf{\large \VII\ Symmetry analysis about the nonlinear optical response}
\end{center}

The nonlinear optical response can be characterized by the second order optical conductivity as 
\begin{equation}
    j^c(\omega_1+\omega_2)=\sum_{a,b}\,\sigma^c_{ab}(\omega_1+\omega_2)E^a(\omega_1)E^b(\omega_2).
    \label{eq:current}
\end{equation}
where $j^{c}(\omega_1+\omega_2)$ denotes the photo current density with frequency $\omega_1+\omega_2$, $E^{a}(\omega_1)$ and $E^{b}(\omega_2)$ denote the electric fields with frequency $\omega_1$ and $\omega_2$, and $a,b,c=x,y$ for 2D systems. $\sigma^{c}_{ab}(\omega_1+\omega_2)$ is the frequency dependent second-order photo conductivity. In this work we consider two kinds of nonlinear optical responses: the second-harmonic generation (SHG), with $\omega_1=\omega_2=\omega$, and the shift-current generation with $\omega_1=-\omega_2=\omega$. In the SHG process, the frequency dependent second-order susceptibility is related to the photo conductivity via $\chi^{c}_{ab}(2\omega)=i\sigma^{c}_{ab}(2\omega)/(2\varepsilon_0\omega)$, where $\omega$ is the frequency of the incident light.
 
The  second-order photo-conductivity in the TBG system can be decomposed into two components: an intrinsic component $\sigma^c_{ab,0}$ that results from the structural $C_{2z}$ symmetry breaking, say, due to hBN alignment;  and another  component that is induced by the order parameter denoted by $\mathbf{\mathcal{N}}$, which can be expressed as $\sum_{d}\sigma^c_{ab,d} \mathcal{N}_d$ where $\mathcal{N}_d$ is the $d$th component of the order parameter. Including both the intrinsic contribution and the order-parameter contribution, the second-order photo conductivitiy can be written as
\begin{equation}
\sigma^c_{ab}(\omega)=\sigma^c_{ab,0}(\omega)+\sum_d\,\sigma^c_{ab,d}(\omega)\mathcal{N}_d.
\end{equation}

An order parameter $\mathbf{\mathcal{N}}$ is transformed by a symmetry operation $\mathit{g}$ as: $\mathcal{N}_d\to \sum_{d'}\gamma(\mathit{g})_{dd'}\,N_{d'}$, where  $\gamma(\mathit{g})_{dd'}$ is the matrix element of the symmetry operator $\mathit{g}$ represented by the order parameter. Now we perform this symmetry transformation on  both sides of Eq.~(\ref{eq:current}),
\begin{equation}
    \sum_{c'}O(\mathit{g})_{cc'}j^{c'}=\sum_{ab}\sum_{a'b'}\sigma^c_{ab,0}O(\mathit{g})_{aa'}E^{a'}O(\mathit{g})_{bb'}E^{b'}+\sum_{ab}\sum_{a'b'}\sum_{d'}\sigma^c_{ab,d}O(\mathit{g})_{aa'}O(\mathit{g})_{bb'}\gamma(\mathit{g})_{dd'}E^{a'}E^{b'}\mathcal{N}_{d'}.
\end{equation}
Here $O(\mathit{g})$ is symmetry operator represented in Cartesian coordinates. From the above equation, we obtain the constraint on the nonlinear photo conductivity $\sigma^{c}_{ab}$ from symmetry $\mathit{g}$:
\begin{align}
\begin{split}
    &\sum_{abc}O^T(\mathit{g})_{c_0c}\sigma^c_{ab,0}O(\mathit{g})_{aa'}O(\mathit{g})_{bb'}=\sigma^{c_0}_{a'b',0},\\
    &\sum_{abcd}O^T(\mathit{g})_{c_0c}\sigma^c_{ab,d}O(\mathit{g})_{aa'}O(\mathit{g})_{bb'}\gamma(\mathit{g})_{dd'}=\sigma^{c_0}_{a'b',d'}.
    \label{eq:symmetry-constraint}
\end{split}
\end{align}
In TBG,  the intrinsic component $\sigma^c_{ab,0}$ vanishes due to $C_{2z}$ symmetry, and  we only need to consider the nonlinear photo conductivities of the various spontaneous symmetry-breaking states induced by  order parameters.

\vspace{12pt}
\begin{center}
\textbf{\large \VII\ A Intervalley coherent states}
\end{center}

First we focus on the IVC states. The nonlinear optical response of a generic IVC ordered state can be generally expressed as
\begin{equation}
    \sigma^c_{ab}=\sigma^c_{ab,x}\mathcal{N}^{IVC}_x+\sigma^c_{ab,y}\mathcal{N}^{IVC}_y,
\end{equation}
where  a generic IVC order has been written as a two-component vector: $\mathbf{\mathcal{N}}^{IVC}=(\tau_x\sigma_{\alpha},\tau_y\sigma_{\alpha})$ ($\alpha=0, x, y, z$), with $\mathbf{\tau}$ and $\mathbf{\sigma}$ denoting Pauli matrices defined in the valley and sublattice space respectively. 
The $C_{2z}=\tau_x\sigma_x$ and $C_{2z}^{\prime}=\tau_zC_{2z}$ symmetry operations are important in determining the nonlinear optical response of the IVC states. Therefore, based on the transformation properties under $C_{2z}$ and $C_{2z}'$ operations, 
IVC states can be divided into two groups: $(\tau_x,\tau_y)\sigma_{y,z}$, and $(\tau_x,\tau_y)\sigma_{0,x}$. In the first group, the $(\tau_x,\tau_y)\sigma_y$  order is known as the K-IVC order, which breaks time-reversal symmetry ($\mathcal{T}$) but preserves a ``Kramers" time-reversal symmetry $\mathcal{T}'=\tau_z\mathcal{T}$ \cite{zaletel-tbg-hf-prx20}; and the $(\tau_x,\tau_y)\sigma_{0,x}$ order is  known as the time-reversal invariant IVC (T-IVC) order, which preserves time-reversal symmetry. 

In the first group of IVC order $(\tau_x,\tau_y)\sigma_{y,z}$, the symmetry representations of $C_{2z}$ and $C_{2z}'$ operations are 
\begin{equation}
    \gamma(C_{2z})=\begin{pmatrix} -1 & 0\\ 0 & 1 \end{pmatrix}, \hspace{6pt}
    \gamma(C_{2z}^\prime)=\begin{pmatrix} 1 & 0\\ 0 & -1 \end{pmatrix}
    \label{eq:ivc1-c2z}
\end{equation}
Plugging Eq.~(\ref{eq:ivc1-c2z}) into Eq.~(\ref{eq:symmetry-constraint}), it follows that
 $C_{2z}$ symmetry only allows the $\sigma^c_{ab,x}$ component,  but $C_{2z}^\prime$ symmetry only allows the $\sigma^c_{ab,y}$ component. Thus all components of nonlinear optical responses are killed by combination of $C_{2z}$ and $C_{2z}^{\prime}$ symmetry operations  for this type of IVC state. For the second group of IVC order $(\tau_x,\tau_y)\sigma_{0,x}$, the symmetry representations of $C_{2z}$ and $C_{2z}'$ operations  become
\begin{equation}
    \gamma(C_{2z})=\begin{pmatrix} 1 & 0\\ 0 & -1 \end{pmatrix},
    \gamma(C_{2z}^\prime)=\begin{pmatrix} -1 & 0\\ 0 & 1 \end{pmatrix}
\end{equation}
It follows that the $C_{2z}$ operation only allows the $\sigma^c_{ab,y}$ component, but $C_{2z}^\prime$ operation only allows the $\sigma^c_{ab,x}$ component  for this group of IVC state. Therefore, the nonlinear optical responses of all the IVC states $(\tau_x,\tau_y)\sigma_{y,z}$ and  $(\tau_x,\tau_y)\sigma_{0,x}$ are vanishing due to the constraints from $C_{2z}$ and $C_{2z}^{\prime}$ symmetries.

\vspace{12pt}
\begin{center}
\textbf{\large \VII\ B Valley polarized states}
\end{center}
Let us continue to discuss the nonlinear optical response for the valley polarized order $\tau_z$. The  symmetry representations of the $\tau_z$ order are given by 
\begin{equation}
    \gamma(C_{3z})=1, \gamma(C_{2y})=-1, \gamma(C_{2x})=1, \gamma(C_{2z})=-1, \gamma(\mathcal{T})=-1 \quad \rm{for}\;\tau_z\;\rm{order\;parameter}.
\end{equation}
The $C_{3z}$ symmetry enforces
\begin{equation}
\begin{split}
    \sigma^x_{xx,0}&=-\sigma^y_{xy,0}=-\sigma^y_{yx,0}=-\sigma^x_{yy,0},\\
    \sigma^x_{xx,z}&=-\sigma^y_{xy,z}=-\sigma^y_{yx,z}=-\sigma^x_{yy,z},\\
    \sigma^y_{xx,0}&=\sigma^x_{xy,0}=\sigma^x_{yx,0}=-\sigma^y_{yy,0},\\
    \sigma^y_{xx,z}&=\sigma^x_{xy,z}=\sigma^x_{yx,z}=-\sigma^y_{yy,z}.
\end{split}
\label{sigma-rest}
\end{equation}
The $C_{2z}$ symmetry enforces $\sigma^c_{ab,0}$ to be vanishing, and $C_{2x}$ enforces $\sigma^y_{xx,z}=0$. Therefore, there are only four symmetry allowed components of nonlinear photo conductivities for the valley polarized state:
\begin{equation}
\sigma^x_{xx,z}=-\sigma^y_{xy,z}=-\sigma^y_{yx,z}=-\sigma^x_{yy,z}\;.
\end{equation}
Clearly, such nonlinear optical response is purely induced by the valley polarization.

\vspace{12pt}
\begin{center}
\textbf{\large \VII\ C Sublattice polarized states}
\end{center}
As for the $\sigma _z$ sublattice polarized order parameter, the symmetry representations are
\begin{equation}
    \gamma(C_{3z})=1, \gamma(C_{2y})=-1, \gamma(C_{2x})=-1, \gamma(C_{2z})=-1, \gamma(\mathcal{T})=1\quad \rm{for}\;\sigma_z\;\rm{order\;parameter}.
\end{equation}
Compared to $\tau_z$ order parameter, the $C_{2x}$ symmetry allows the non-zero $\sigma^y_{xx,z}$. Thus, the nonlinear optical response with the $\sigma _z$ order parameter becomes 
\begin{align}
&\sigma^x_{xx,z}=-\sigma^y_{xy,z}=-\sigma^y_{yx,z}=-\sigma^x_{yy,z}\;\nn
&\sigma^y_{xx,z}=\sigma^x_{xy,z}=\sigma^x_{yx,z}=-\sigma^y_{yy,z}
\end{align}

\vspace{12pt}
\begin{center}
\textbf{\large \VII\ D Nematic ordered states}
\end{center}
Another type of order parameter which exhibit non-vanishing nonlinear optical response is a kind of nematic order that is predicted as the candidate state of twisted bilayer-monolayer and twisted double-bilayer graphene systems \cite{zhangsh-tbmg-arxiv21}, which consists of two components $\mathbf{\mathcal{N}}^{\textrm{nem}}=(\tau_z\sigma_x,\sigma_y)$. Then the nonlinear photo conductivity in such a state is expressed as $\sigma_{ab}^c=\sigma_{ab,x}^c\mathbf{\mathcal{N}}^{\textrm{nem}}_x+\sigma_{ab,y}^c\mathbf{\mathcal{N}}^{\textrm{nem}}_y$.  The symmetry representations for $\mathbf{\mathcal{N}}^{\textrm{nem}}$ are expressed as
\begin{equation}
    \gamma(C_{3z})=\begin{pmatrix} -\frac{1}{2} & \frac{\sqrt{3}}{2}\\ -\frac{\sqrt{3}}{2} & -\frac{1}{2} \end{pmatrix},
    \gamma(C_{2z})=\begin{pmatrix} -1 & 0\\ 0 & -1 \end{pmatrix},
    \gamma(C_{2z}^\prime)=\begin{pmatrix} -1 & 0\\ 0 & -1 \end{pmatrix},
    \gamma(C_{2x})=\begin{pmatrix} 1 & 0\\ 0 & -1 \end{pmatrix}.
\end{equation}
The $C_{2z}$ and $C_{2z}^{\prime}$ operations allow all components of nonlinear photo conductivities, but the $C_{2x}$ operation kills the $\sigma_{xx,y}^x$, $\sigma_{xy,y}^y$, $\sigma_{yx,y}^y$, $\sigma_{yy,y}^x$, $\sigma_{xx,x}^y$, $ \sigma_{xy,x}^x$, $\sigma_{yx,x}^x$ and $\sigma_{yy,x}^y$ components. The $C_{3z}$ operation further requires the remaining non-vanishing components to satisfy the following conditions for the nematic order $\mathbf{\mathcal{N}}^{\textrm{nem}}=(\tau_z\sigma_x,\sigma_y)$
\begin{align}
&\sigma^{x}_{xx,x}=\sigma_{xy,x}^{y}+\sigma_{yx,x}^{y}+\sigma_{yy,x}^{x}\;\nn
&\sigma^{y}_{yy,y}=\sigma_{xx,y}^{y}+\sigma_{xy,y}^{x}+\sigma_{yx,y}^{x}\;\nn
&\sigma^{x}_{xx,x}=-\sigma^{y}_{yy,y}\;\nn
&\sigma_{xy,x}^{y}=-\sigma_{yx,y}^{x}\;\nn
&\sigma_{yx,x}^{y}=-\sigma_{xy,y}^{x}\;\nn
&\sigma_{yy,x}^{x}=-\sigma_{xx,y}^{y}\;,
\end{align}

As for another kind of nematic order $(\sigma_x,\tau_z\sigma_y)$, the symmetry representation of $C_{2z}$ operation is
\begin{equation}
\gamma(C_{2z})=\begin{pmatrix} 1 & 0\\ 0 & 1 \end{pmatrix},
\end{equation}
which kills all components of nonlinear optical response in this ordered state. Similarly, the $C_{2z}$ operation also forbids all components of nonlinear optical response in the $\tau_z\sigma_z$ order. In summary, there are only three types of orders $\tau_z$, ($\tau_z\sigma_x$, $\sigma_y$) and $\sigma_z$ orders which can have non-vanishing nonlinear optical responses, which are explicitly shown in Table~\ref{table:nonlinear}. 

\vspace{12pt}
\begin{center}
\textbf{\large \VII\ E Strain effects}
\end{center}
We continue to discuss the strain effects on the nonlinear optical responses. We note that at the charge neutrality point, strain may drive a transition from  the K-IVC correlated insulator state to a non-interacting semi-metallic state \cite{bultinck-tbg-strain-prl21}. Typically  the TBG system would preserves $C_{2z}$ symmetry under in-plane strain, thus the nonlinear optical response of the non-interacting semi-metallic phase would be vanishing. On the other hand, a pure K-IVC state also has vanishing nonlinear optical response as discussed above. Therefore, if the correlated insulator at CNP is a pure K-IVC state, and if strain does not break $C_{2z}$ symmetry of the moir\'e superlattice, the in-plane components of the second-order susceptibility would remain vanishing through transition from the K-IVC state to the semi-metallic state at charge neutrality point.

\begin{figure}[!htbp]
\includegraphics[width=0.4\textwidth]{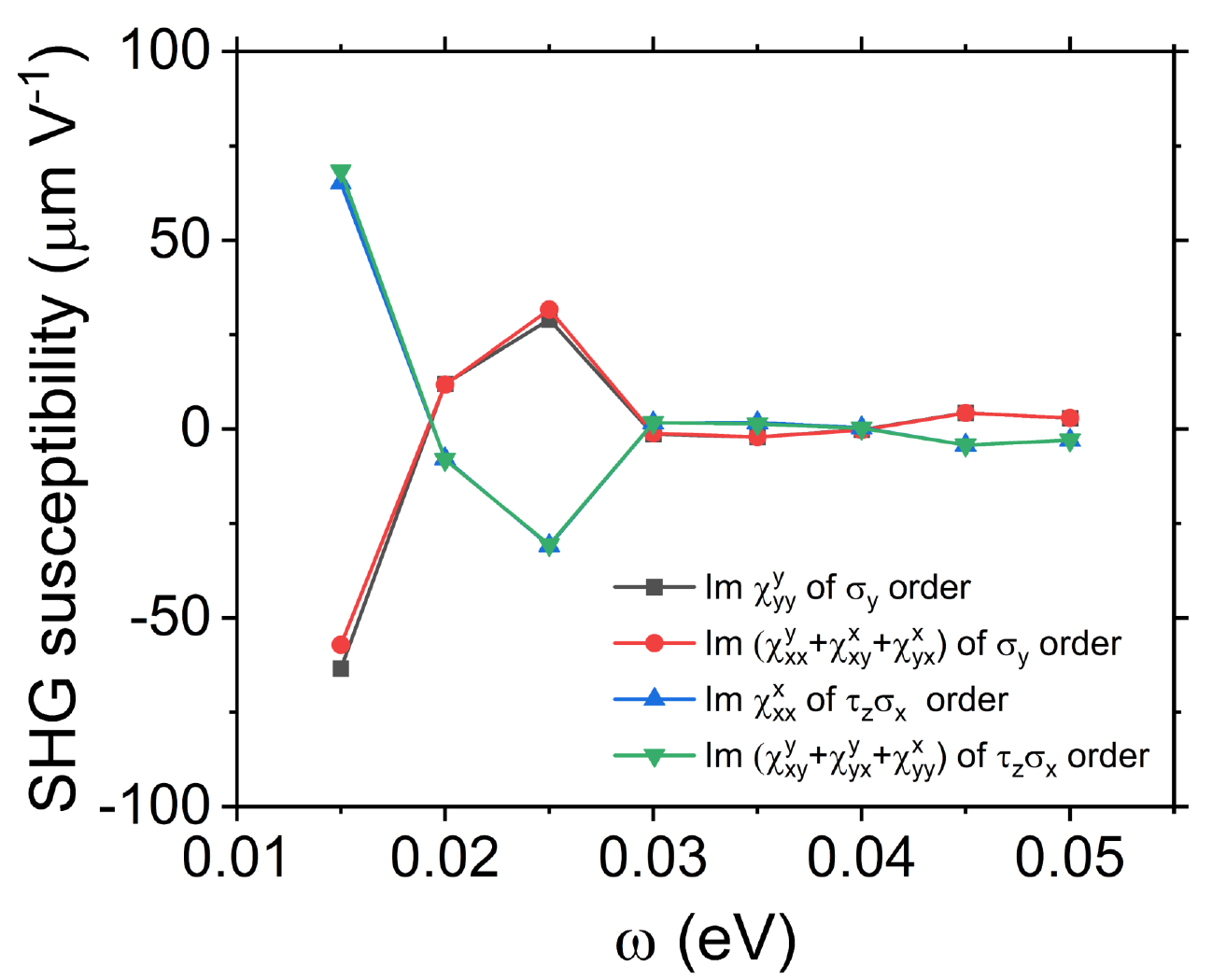}
\caption{~\label{resfig2} The SHG susceptibility with constant nematic order in the TBG. We input a constant, $\kt$-independent order parameter with amplitude of 1\,meV in $\tau_z\sigma_x$ or $\sigma_y$ ordered state in the calculations.}
\end{figure}

At other integer or fractional filling factors, the ground states may be intervalley coherent (IVC) state, valley polarized state, or the incommensurate Kekul\'e spiral (IKS) state. As discussed above, the nonlinear optical response of the IVC state vanishes, but the valley polarized state has non-vanishing nonlinear optical response. In what follows we will show that the IKS state may also exhibit non-linear optical properties. To be specific, it has been proposed by Kwan \textit{et al.} \cite{zaletel-tbg-hf-prx20} that, if there exists some strain in the TBG system, the ground state would be the IKS state at non-zero fillings. The order parameter in the IKS phase can be defined as $(1+\bm{n}_{\kt}\cdot\bm{\gamma})(1+\bm{m}_{\kt}\cdot\bm{\eta})$, where $\gamma\!=\!(\sigma _x, \tau _z\sigma _y, \tau_z\sigma_z)$ and $\eta\!= \!(\tau_x\sigma_x,\tau_y\sigma_x,\tau_z)$, both satisfy the $\rm{SU}(2)$ Lie algebra. 
It can be noted that an equivalent definition of $\mathbf{\gamma}$: $\mathbf{\gamma}=(\tau_z\sigma_x,\sigma_y,\tau_z\sigma_z)$ generate the same order parameters in the IKS state. The order parameters $(\tau_z\sigma_x,\sigma_y)$ in the IKS state are dubbed as ``nematic orders" in the previous subsection, which break $C_{2z}$ symmetry and allow for nonlinear optical response as discussed in Sec.~\VII\ D.  In Fig.~\ref{resfig2} we show the calculated SHG susceptibility in the symmetry-breaking state with the ``nematic order" $(\tau_z\sigma_x,\sigma_y)$. Clearly we see remarkable SHG responses, and calculated the susceptibility tensor elements are compatible with the symmetry analysis discussed in Sec.~~\VII\ D. If the ground state at some given filling is an IVC state (say at 7/2 filling from our calculations), and strain may drive the system to transit from the IVC state to the IKS state, then such a strain-induced transition at fixed filling factor can be experimentally characterized by nonlinear optical response.



\begin{table}[!t]
 \caption{Three types of ordered states with non-vanishing nonlinear optical responses}
 \label{unitcell}
 \centering
\begin{tabular}{cccc} 
 \hline
  order parameters &  symmetry-allowed nonlinear  conductivities \\
  \hline
  $\tau_z$  &    $\sigma_{xx}^{x}=-\sigma_{xy}^{y}=-\sigma_{yx}^{y}=-\sigma^{x}_{yy}$ \\
 \hline
  $(\tau_z\sigma_x, \sigma_y)$  & $\sigma^{x}_{xx,x}=\sigma_{xy,x}^{y}+\sigma_{yx,x}^{y}+\sigma_{yy,x}^{x}$,  \hspace{3pt} $\sigma^{y}_{yy,y}=\sigma_{xx,y}^{y}+\sigma_{xy,y}^{x}+\sigma_{yx,y}^{x}$, \\
   & $\sigma^{x}_{xx,x}=-\sigma^{y}_{yy,y}$, \hspace{3pt} $\sigma_{xy,x}^{y}=-\sigma_{yx,y}^{x}$, \hspace{3pt} $\sigma_{yx,x}^{y}=-\sigma_{xy,y}^{x}$, \hspace{3pt} $\sigma_{yy,x}^{x}=-\sigma_{xx,y}^{y}$ \\
  \hline
  $\sigma_z$ &  $\sigma_{xx}^{x}=-\sigma_{xy}^{y}=-\sigma_{yx}^{y}=-\sigma^{x}_{yy}$,  \\
   &  $\sigma_{xx}^{y}=\sigma_{xy}^{x}=\sigma_{yx}^{x}=-\sigma^{y}_{yy}$\\
 \hline
 \label{table:nonlinear}
\end{tabular}
\end{table}

\vspace{12pt}
\begin{center}
\textbf{\large \VII\ F Experimental outputs}
\end{center}
Now, we discuss about what kind of information can we obtain from the outputs of second harmonic generation measurements. According to previous studies, the candidate ground states at different filling factors of magic-angle TBG include: the valley polarized state, the K-IVC state, the non-interacting semi-metallic state, and the IKS state. These states would have distinct nonlinear optical responses, and  each type of the correlated states can be uniquely and unambiguously determined using SHG measurements combined with linear transport measurements. To be specific, 
\begin{itemize}
\item If the ground state is a valley polarized state, then there are only four non-zero components in the SHG susceptibility tensor: $\chi_{xx}^{x}=-\chi_{xy}^{y}=-\chi_{yx}^{y}=-\chi^{x}_{yy}$.
\item If the ground state is a K-IVC  insulator state or a non-interacting semi-metallic state (e.g., at the CNP), then the SHG response vanishes. Linear transport measurements can further help to distinguish between the K-IVC insulator state and the semi-metallic state.
\item If the ground state is an IKS state which involves the order parameter $(\tau_z\sigma_x,\sigma_y)$, then there would be eight non-zero components in the SHG susceptibility tensor, as given in Table.~\II.\\
\end{itemize}



\vspace{12pt}
\begin{center}
\textbf{\large \VIII\ Microscopic expressions for the nonlinear photo conductivities}
\end{center}
The nonlinear photo conductivity can be derived based on time-dependent second-order perturbation theory. Specifically, we start from the Liouville-von Neumann equation in the interaction picture
\begin{equation}
\frac{d\hat{\rho}_I(t)}{dt}=\frac{-i}{\hbar}\,[\hat{H}^{\textrm{ext}}_I(t),\hat{\rho}_I(t)]\;
\end{equation}
where $\hat{O}_I=e^{i\hat{H}_0t/\hbar}\,\hat{O}\,e^{-i\hat{H}_0 t/\hbar}$. The external vector potential $\mathbf{A}(\mathbf{r},t)$ couples to the current density $\hat{\mathbf{j}}(\mathbf{r},t)$ as
\begin{equation}
\hat{H}^{\textrm{ext}}_I(t)=-\int d\mathbf{r}\,\hat{\mathbf{j}}_{I}(\mathbf{r},t)\cdot\mathbf{A}(\mathbf{r},t)\;,
\end{equation}
and the electric field $\mathbf{E}(\mathbf{r},t)=-\partial\mathbf{A}(\mathbf{r},t)/\partial t$.
We expand the density matrix $\hat{\rho}_I(t)$ to the second order in $\hat{H}^{\textrm{ext}}_I(t)$,  i.e., $\hat{\rho}_I(t)=\hat{\rho}_0+\delta\hat{\rho}^{(1)}_I(t)+\delta\hat{\rho}^{(2)}_I(t)+...$, and finds that
\begin{equation}
\delta\hat{\rho}^{(2)}_I(t)=(\frac{-i}{\hbar})^2\int_{-\infty}^{t}\,dt'\,[\hat{H}^{\textrm{ext}}_I(t'), \int_{-\infty}^{t'}dt''\,[\hat{H}^{\textrm{ext}}_I(t''),\hat{\rho}_0]\,]\;.
\end{equation}
Then we can calculate the expectation value of the current density up to the second order response of the external fields, which gives us the second-order photo conductivity.
In particular, for the shift-current generation in response to linearly polarized light, the nonlinear conductivity with incident light frequency $\omega$ is expressed as \cite{baltz-prb81,zhang-wsm-prb18,jpliu-tbg-optics-npj20}
\begin{equation}
\sigma ^c_{ab}(\omega,0)=\frac{e^3}{\omega^2\,d}\int\frac{d^2k}{(2\pi)^2}\sum_{nml}\sum_{\Omega=\pm\omega}\,\textrm{Re}\,[\,\frac{(f_{l\mathbf{k}}-f_{n\mathbf{k}})\,v^a_{nl}v^b_{lm}v^c_{mn}}{(E_{n\mathbf{k}}-E_{l\mathbf{k}}+\hbar\Omega-i\delta)(E_{n\mathbf{k}}-E_{m\mathbf{k}}-i\delta)}\,].
\label{eq:shift-current}
\end{equation}
where $d\approx 3.35\,\angstrom$ is the thickness of TBG, $n, m, l$ are the band indices, $v^a_{nl}=\langle u_{n\mathbf{k}} | \partial _{k_a}H_{\mathbf{k}} | u_{l\mathbf{k}} \rangle / \hbar$ is  matrix element of the velocity operator, $f_{n\mathbf{k}}$ is the Fermi-Dirac distribution function with respect to the band energy $E_{n\mathbf{k}}$, and $\delta=\hbar/\tau$ is a small smearing factor arising from the finite quasi-particle lifetime $\tau$ of the photo-excited electrons, which is set to $\delta=0.5\,$meV in our calculations. 

In the valley polarized state of TBG, the shift current is solely induced by the time-reversal breaking order parameter $\tau_z$, and such shift current generated in  a $\mathcal{T}$-broken state in response to linearly polarized light is also dubbed as  ``injection current" in a recent study about magnetization-induced nonlinear optical response in bilayer antiferromagnetic CrI$_3$ \cite{zhang-cri3-nc19}. This is because as a result of the $\mathcal{T}$ symmetry breaking, the real part of the velocity matrix elements in Eq.~(\ref{eq:shift-current})   $\textrm{Re}[v^a_{nl}v^b_{lm}v^c_{mn}]$ no longer cancel each other for opposite $\mathbf{k}$ points, as opposite to the $\mathcal{T}$-invariant case, in which $\mathcal{T}$ symmetry requires $v^a_{nl}(\mathbf{k})v^b_{lm}(\mathbf{k})v^c_{mn}(\mathbf{k})=-(v^a_{nl}(\mathbf{-k})v^b_{lm}(\mathbf{-k})v^c_{mn}(\mathbf{-k}))^{*}$. Therefore, the real part of the energy denominator in Eq.~(\ref{eq:shift-current}) combined with the real part of the velocity matrix elements would make significant contributions to the shift current response. This $\mathcal{T}$-breaking contribution to the shift current is dominated by a two-band process with $m=n$, which is exactly proportional to $1/\delta=\tau/\hbar$, i.e. proportional to the quasi-particle lifetime $\tau$ \cite{zhang-cri3-nc19}, which is reminiscent of the injection current in $\mathcal{T}$-invariant system driven by circularly polarized light. Therefore, the $\mathcal{T}$-breaking contribution to the shift current in response to linearly polarized light is also called ``injection current" or ``injection-like current" in literatures \cite{zhang-cri3-nc19,yan-prr20}.

Following the same procedure, one obtains the nonlinear photo conductivity for  second harmonic generation 
\begin{equation}
\begin{split}
\sigma ^c_{ab}(\omega;2\omega)&=-\frac{e^3}{\omega^2\,d}\int\frac{d^2k}{(2\pi)^2}\sum_{nml}(f_{n\mathbf{k}}-f_{l\mathbf{k}})\frac{v^a_{nm}v^c_{ml}v^b_{ln}}{(E_{l\mathbf{k}}-E_{n\mathbf{k}}-\hbar\omega-i\delta)(E_{l\mathbf{k}}-E_{m\mathbf{k}}-2\hbar\omega-i\delta)}\\
&+\frac{e^3}{\omega^2\,d}\int\frac{d^2k}{(2\pi)^2}\sum_{nml}(f_{n\mathbf{k}}-f_{l\mathbf{k}})\frac{v^c_{nm}v^a_{ml}v^b_{ln}}{(E_{l\mathbf{k}}-E_{n\mathbf{k}}-\hbar\omega-i\delta)(E_{m\mathbf{k}}-E_{n\mathbf{k}}-2\hbar\omega-i\delta)}.
\end{split}
\label{eq:shg}
\end{equation}
and the SHG susceptibility $\chi^{c}_{ab}=i\sigma^{c}_{ab}/(2\epsilon_0\omega)$. Note that after permuting band indices, Eq.~(\ref{eq:shg}) is the same as that of Eq.~(S7) in Supplementary Information of Ref.~\onlinecite{gao-prl20}. In Fig.~4(a) of main text, we have presented the imaginary part of the SHG susceptibility for different types of ordered states.


\end{document}